\documentclass{aa}
\usepackage{epsfig}
\usepackage{natbib}
\bibpunct{(}{)}{;}{a}{}{,}

\newbox\grsign \setbox\grsign=\hbox{$>$} \newdimen\grdimen \grdimen=\ht\grsign
\newbox\simlessbox \newbox\simgreatbox \newbox\simpropbox \newbox\wtildebox 
\setbox\simgreatbox=\hbox{\raise.5ex\hbox{$>$}\llap
     {\lower.5ex\hbox{$\sim$}}}\ht1=\grdimen\dp1=0pt
\setbox\simlessbox=\hbox{\raise.5ex\hbox{$<$}\llap
     {\lower.5ex\hbox{$\sim$}}}\ht2=\grdimen\dp2=0pt

\newcommand{\be}{\mbox{\begin{equation}}}
\newcommand{\ee}{\mbox{\end{equation}}}

\newcommand{\tdis}{\mbox{$t_{\rm dis}^{\rm total}$}}

\newcommand{\Cref}{\mbox{$m_{\rm ref}$}}

\newcommand{\msun}{\mbox{M$_\odot$}}

\renewcommand{\d}{{\rm d}} 
\newcommand{\mc}{\mbox{$M_{\rm cl}$}}
\newcommand{\mci}{\mbox{$M_{\rm cl,i}$}}
\newcommand{\mct}{\mbox{$M_{\rm cl}(t)$}}
\newcommand{\mcli}{\mbox{$M_{\rm cl,i}$}}

\begin{document}

\title{Dissolution is the solution: on the reduced mass-to-light ratios of Galactic globular clusters
}

\author{J.~M.~Diederik~Kruijssen \inst{1,2} and
        Steffen~Mieske \inst{3}}

\institute { 
                 {\inst{1}Astronomical Institute, Utrecht University, 
                 PO Box 80000, 3508TA Utrecht, The Netherlands\\
                 (e-mail: {\tt kruijssen@astro.uu.nl})}\\
                 {\inst{2}Sterrewacht Leiden, Leiden University,
                 PO Box 9513, 2300RA Leiden, The Netherlands}\\
                 {\inst{3}European Southern Observatory, Alonso de Cordova 3107, Vitacura, Santiago, Chile\\
                 (e-mail: {\tt smieske@eso.org})}
                 }

\date{Received date ; accepted date}


\abstract{The observed dynamical mass-to-light ($M/L$) ratios of globular clusters (GCs) are systematically lower than the value expected from `canonical' simple stellar population models, which do not account for dynamical effects such as the preferential loss of low-mass stars due to energy equipartition. It was recently shown that low-mass star depletion can qualitatively explain this discrepancy {{for globular clusters}} in several galaxies.}
{To verify whether low-mass star depletion is indeed the driving mechanism behind the $M/L$ decrease, we aim to predict the $M/L_V$ ratios of individual GCs for which orbital parameters and dynamical $V$-band mass-to-light ratios $M/L_V$ are known. There is a sample of 24 Galactic GCs for which this is possible.}
{We use the {{\tt SPACE}} cluster models, which include dynamical dissolution, low-mass star depletion, stellar evolution, stellar remnants and various metallicities. We derive the dissolution timescales due to {{two-body relaxation and disc shocking}} from the orbital parameters of our GC sample and use these to predict the $M/L_V$ ratios of the individual GCs. To verify our findings, we also predict the slopes of their low-mass stellar mass functions.}
{The computed dissolution timescales are in good agreement with earlier empirical studies. The predicted $M/L_V$ are in $1\sigma$ agreement with the observations for 12 out of 24 GCs. The discrepancy for the other GCs probably arises because our predictions give global $M/L$ ratios, while the observations represent extrapolated central values that are different from global ones in case of mass segregation and a long dissolution timescale. GCs in our sample which likely have dissimilar global and central $M/L$ ratios can be excluded by imposing limits on the dissolution timescale and King parameter. For the remaining GCs, the observed and predicted average $M/L_V$ are 78$^{+9}_{-11}$\% and $78\pm2$\% of the canonically expected values, while for the entire sample the values are 74$^{+6}_{-7}$\% and $85\pm1$\%. The predicted correlation between the slope of the low-mass stellar mass function and $M/L_V$ drop is found to be qualitatively consistent with observed mass function slopes.}
{The dissolution timescales of Galactic GCs are such that the $\sim 20$\% gap between canonically expected and observed $M/L_V$ ratios is bridged by accounting for the preferential loss of low-mass stars, also when considering individual clusters. It is concluded that the variation of $M/L$ ratio due to dissolution and low-mass star depletion is a plausible explanation for the discrepancy between the observed and canonically expected $M/L$ ratios of GCs.}

\keywords{
Galaxy: globular clusters: general --
Galaxy: stellar content --
galaxies: star clusters
}

\authorrunning{J.~M.~D.~Kruijssen and S.~Mieske}
\titlerunning{On the reduced mass-to-light ratios of Galactic globular clusters}

\maketitle


\section{Introduction} \label{sec:intro}
The topic of dynamical mass-to-light ($M/L$) ratios of old
compact stellar systems has attracted increasing attention during recent years
\citep{mclaughlin05,hasegan05,rejkuba07,hilker07,evstigneeva07,dabringhausen08,mieske08,kruijssen08b,baumgardt08,chilingarian08,forbes08}. The outcome of these studies can be summarised as follows:
\begin{itemize}

\item For the mass regime of ultra-compact dwarf galaxies (UCDs, $M \geq 2 \times 10^6~\msun$), dynamical $M/L$ ratios tend to be some 50\% above predictions from stellar population models \citep{dabringhausen08, mieske08, forbes08}.

\item For the mass regime of globular clusters (GCs, $M \leq 2 \times 10^6~\msun$), dynamical $M/L$ ratios tend to be some 25\% below predictions from simple stellar population (SSP) models that assume a canonical IMF \citep{rejkuba07,kruijssen08b,kruijssen08,mieske08}.

\item As a consequence, the dynamical $M/L$ ratios of UCDs are on average about twice as large as those of GCs, at comparable metallicities.

\end{itemize}

Regarding GCs, a viable solution to obtain {\it lower} $M/L$ ratios is
a deficit in low-mass stars with respect to a canonical initial mass function (IMF) \citep{kroupa01}. {This is known to arise naturally from two-body relaxation in star clusters, which causes a depletion of low-mass stars \citep{vesperini97b,baumgardt03}.} In \citet{kruijssen08b} it has been studied how the
preferential loss of low-mass stars due to dynamical evolution of a
star cluster in a tidal field reduces the $M/L$ ratios of star
clusters. There, we constrained the ranges of dissolution
timescales necessary for this loss of low-mass stars to quantitatively
account for the drop of $M/L$ observed for GCs. In the
case of the Galactic GC system, it was found that dissolution
timescales in the range $t_0=0.6$---$\geq$20~Myr (corresponding to
total disruption times of $\tdis\approx3$---100~Gyr for a $10^6~\msun$
cluster) are required to explain the observed $M/L$ ratio
decline. {It was also shown that the $M/L$ ratio decrease is strongest for low-mass GCs, which explains the observed correlation of increasing $M/L$ ratio with mass discovered by \citet{mandushev91}. \citet{kruijssen08b} concluded that the scatter around this relation is caused by spreads in metallicity and dissolution timescale.}

As noted already in \citet[hereafter KL08]{kruijssen08}, the next step
is to apply these analytical cluster models including preferential
loss of low-mass stars to {\it individual} clusters. {This would then account for variations in metallicity and dissolution timescale.} Such a study will
naturally be restricted to GCs with measured $M/L$ ratios for which realistic
estimates of their individual dissolution timescale are available from
information on their actual orbit within the Milky Way potential. With
the database of individual dissolution time scales at hand, the loss
of low-mass stars can be quantified according to the prescriptions of
\citet{kruijssen08b} and KL08, leading to predictions for the drop of
$M/L$ for {\it individual} GCs. Those predictions are to be contrasted
with the actual observed $M/L$ ratios of these GCs. This will allow us
to quantitatively test the hypothesis that the loss of low-mass stars is
responsible for the too low $M/L$ ratios of GCs, and hence also partially for
the discrepancy of $M/L$ between GCs and UCDs.

Previous studies assessing the preferential loss of low-mass stars in Galactic GCs focus both on observations \citep[e.g.,][]{demarchi07} and theory \citep[e.g.,][]{baumgardt08b}. In \citet{demarchi07}, the slopes of the stellar mass functions in GCs are measured for stars between 0.3 and 0.8~$\msun$, thereby directly reflecting possible low-mass star depletion. The study by \citet{baumgardt08b} predicts the same slopes using $N$-body models and different degrees of mass segregation, assuming {dissolution by two-body relaxation}. The aforementioned papers both do not consider the $M/L$ ratios of the GCs in question.

\begin{table*}[tb]\centering
\begin{tabular}{c |c c c c c c}
  \hline \multicolumn{7}{c}{Cluster properties} \\
  \hline \hline NGC & $\log{M}^\star$ (\msun) & $(M/L_V)_{\rm obs}^\star$ (\msun~L$_\odot^{-1}$) & [Fe/H]$^{\dagger,\diamond}$ & $R_{\rm gc}^\dagger$ (kpc) & $W_0^\star$ & $(M/L_V)_{\rm can}$ (\msun~L$_\odot^{-1}$)  \\ \hline
    104         &  5.804$_{-0.193}^{+0.157}$         &  1.33$_{-0.59}^{+0.48}$     &  -0.76       &  7.4          &   8.6$_{-0.1}^{+0.1}$  &   $2.68\pm 0.25$         \\
    288         &  4.892$_{-0.198}^{+0.162}$         &  2.15$_{-0.98}^{+0.80}$     &  -1.24       &  12.0        &   4.8$_{-0.2}^{+0.2}$   &  $2.20\pm 0.08$          \\
  1851         &  5.407$_{-0.192}^{+0.156}$         &  1.61$_{-0.71}^{+0.58}$     &  -1.22       &  16.7        &   8.1$_{-0.2}^{+0.1}$  &  $2.21\pm 0.09$         \\
  1904         &  4.984$_{-0.195}^{+0.157}$         &  1.16$_{-0.52}^{+0.42}$     &  -1.57       &  18.8        &   7.5$_{-0.1}^{+0.1}$  &  $2.08\pm 0.04$         \\
  4147         &  4.394$_{-0.202}^{+0.159}$         &  1.01$_{-0.47}^{+0.37}$     &  -1.83       &  21.3        &   8.0$_{-0.1}^{+0.2}$  &  $2.03\pm 0.02$         \\
  4590         &  4.644$_{-0.194}^{+0.156}$         &  0.92$_{-0.41}^{+0.33}$     &  -2.06       &  10.1        &   6.6$_{-0.1}^{+0.1}$  &  $2.00\pm 0.01$         \\
  5139         &  6.503$_{-0.159}^{+0.200}$         &  2.54$_{-0.93}^{+1.17}$     &  -1.29       &  6.4          &   6.2$_{-0.2}^{+0.1}$   &  $2.18\pm 0.07$          \\
  5272         &  5.443$_{-0.197}^{+0.156}$         &  1.39$_{-0.63}^{+0.50}$     &  -1.57       &  12.2        &   8.2$_{-0.1}^{+0.1}$  &  $2.08\pm 0.04$         \\
  5466         &  4.687$_{-0.200}^{+0.162}$         &  1.61$_{-0.74}^{+0.60}$     &  -2.22       &  16.2        &   4.2$_{-0.2}^{+0.2}$  &  $1.99\pm 0.01$         \\
  5904         &  5.252$_{-0.195}^{+0.156}$         &  0.78$_{-0.35}^{+0.28}$     &  -1.27       &  6.2          &   7.6$_{-0.1}^{+0.1}$   &  $2.19\pm 0.08$          \\
  6093         &  5.597$_{-0.205}^{+0.161}$         &  2.67$_{-1.26}^{+0.99}$     &  -1.75       &  3.8          &   7.5$_{-0.1}^{+0.1}$   &  $2.04\pm 0.03$          \\
  6121         &  4.864$_{-0.243}^{+0.178}$         &  1.27$_{-0.71}^{+0.52}$     &  -1.20       &  5.9          &   7.4$_{-0.1}^{+0.1}$   &  $2.22\pm 0.09$           \\
  6171         &  4.922$_{-0.241}^{+0.172}$         &  2.20$_{-1.22}^{+0.87}$     &  -1.04       &  3.3          &   7.0$_{-0.2}^{+0.1}$   &  $2.34\pm 0.13$          \\
  6205         &  5.469$_{-0.201}^{+0.158}$         &  1.51$_{-0.70}^{+0.55}$     &  -1.54       &  8.7          &   7.0$_{-0.1}^{+0.1}$   &  $2.08\pm 0.04$          \\
  6218         &  4.918$_{-0.206}^{+0.157}$         &  1.77$_{-0.84}^{+0.64}$     &  -1.48       &  4.5          &   6.1$_{-0.2}^{+0.1}$   &  $2.10\pm 0.05$          \\
  6254         &  5.234$_{-0.223}^{+0.169}$         &  2.16$_{-1.11}^{+0.84}$     &  -1.52       &  4.6          &   6.5$_{-0.1}^{+0.1}$   &  $2.09\pm 0.04$          \\
  6341         &  5.084$_{-0.202}^{+0.163}$         &  0.88$_{-0.41}^{+0.33}$     &  -2.28       &  9.6          &   7.5$_{-0.1}^{+0.1}$   &  $1.99\pm 0.01$          \\
  6362         &  4.764$_{-0.191}^{+0.161}$         &  1.16$_{-0.51}^{+0.43}$     &  -0.95       &  5.1          &   5.3$_{-0.2}^{+0.3}$   &  $2.42\pm 0.16$          \\
  6656         &  5.606$_{-0.241}^{+0.172}$         &  2.07$_{-1.15}^{+0.82}$     &  -1.64       &  4.9          &   6.5$_{-0.2}^{+0.1}$   &  $2.06\pm 0.03$          \\
  6712         &  4.906$_{-0.241}^{+0.175}$         &  0.99$_{-0.55}^{+0.40}$     &  -1.01       &  3.5          &   5.1$_{-0.4}^{+0.4}$   &  $2.37\pm 0.14$          \\
  6779         &  4.911$_{-0.165}^{+0.207}$         &  1.05$_{-0.40}^{+0.50}$     &  -1.94       &  9.7          &   6.5$_{-0.2}^{+0.3}$   &  $2.01\pm 0.02$          \\
  6809         &  5.219$_{-0.067}^{+0.054}$         &  3.23$_{-0.50}^{+0.40}$     &  -1.81       &  3.9          &   4.5$_{-0.1}^{+0.2}$   &  $2.03\pm 0.02$          \\
  6934         &  5.099$_{-0.193}^{+0.155}$         &  1.51$_{-0.67}^{+0.54}$     &  -1.54       &  12.8        &   7.0$_{-0.2}^{+0.1}$  &  $2.08\pm 0.04$         \\
  7089         &  5.561$_{-0.195}^{+0.160}$         &  0.98$_{-0.44}^{+0.36}$     &  -1.62       &  10.4        &   7.2$_{-0.1}^{+0.1}$  &  $2.06\pm 0.03$         \\\hline
\end{tabular} 
\caption[]{\label{tab:obs}
     Observed properties for the cluster sample. Consecutive columns list the cluster NGC number, logarithm of the present-day cluster mass $M$, observed $V$-band mass-to-light ratio $(M/L_V)_{\rm obs}$, metallicity [Fe/H], galactocentric radius $R_{\rm gc}$, King parameter $W_0$ and canonically expected $V$-band mass-to-light ratio $(M/L_V)_{\rm can}$. Masses and observed $M/L_V$ ratios {are dynamical values}. For all clusters, the standard error in [Fe/H] is assumed to be 0.15 when computing the error propagation (see Sect.~\ref{sec:t0}), which represents a conservative accuracy estimate \citep[see, e.g.,][]{carretta97}. These errors determine the uncertainty on $(M/L_V)_{\rm can}$ in the last column, since the canonical $M/L$ ratio only depends on metallicity.
     \\ $^\star$From \citet{mclaughlin05}.
     \\ $^\dagger$From \citet{harris96}. {The $R_{\rm gc}$ values used to compute the orbits in \citet{dinescu99} are from Zinn, private communication. In extreme cases this may cause a small disagreement between the galactocentric radius quoted here and the apogalactic distance predicted by \citet{dinescu99} (see Table~\ref{tab:orbits}).}
     \\ $^\diamond$The value for NGC~5139 ($\omega$Cen) is derived from \citet{bedin04}.
    }
\end{table*}
In this study, the reference sample for dynamical $M/L$ ratios of
Galactic GCs is that of \citet{mclaughlin05}, obtained by the fitting of
{single-mass} King profiles. It contains data for 38 GCs. Only a subsample
can be used for our analysis, namely those clusters for which accurate
proper motions and radial velocities are measured and can be
translated to orbital parameters. Next to the destruction rates
due to {two-body relaxation}, also the influence of {disc
shocking} on the cluster dissolution needs to be taken into account. {Both have to be derived from the orbital parameters.} Several studies in which orbital information is derived and used to
compute destruction rates are available in the literature
\citep{gnedin97,dinescu99,allen06,allen08}, all with certain benefits
and trade-offs. Specifically, \citet{gnedin97} assign statistically
sampled orbits conforming to the bulk motion of the GC
system in an axisymmetric potential to derive destruction rates of 119
globular clusters. \citet{dinescu99} use proper motions and
radial velocities to compute the orbits and destruction rates of 38
clusters in two axisymmetric potentials
\citep{paczynski90,johnston95}. \citet{allen06,allen08} follow the
same procedure, but consider both axisymmetric and barred potentials
\citep[respectively]{allen91,pichardo04} for 54 globular
clusters. They do not find a significant deviation between the results
for both potentials. However, they do note that their calculated
destruction rates are multiple orders of magnitude smaller than others
in literature and recommend combining their orbital information with
the more rigorous Fokker-Planck approach used by \citet{gnedin97} to
derive the destruction rates.

We choose to adopt the orbital data and destruction rate due to disc shocking from \citet{dinescu99}. Our study cannot be based on statistically assigned orbits but requires the actual orbits of individual clusters, thus excluding the estimated orbits from \citet{gnedin97}. In addition, the \citet{dinescu99} destruction rates seem to be in better agreement with the observations than those from \citet{allen06,allen08}.

\begin{figure*}[t]
\resizebox{\hsize}{!}{\includegraphics[width=17cm]{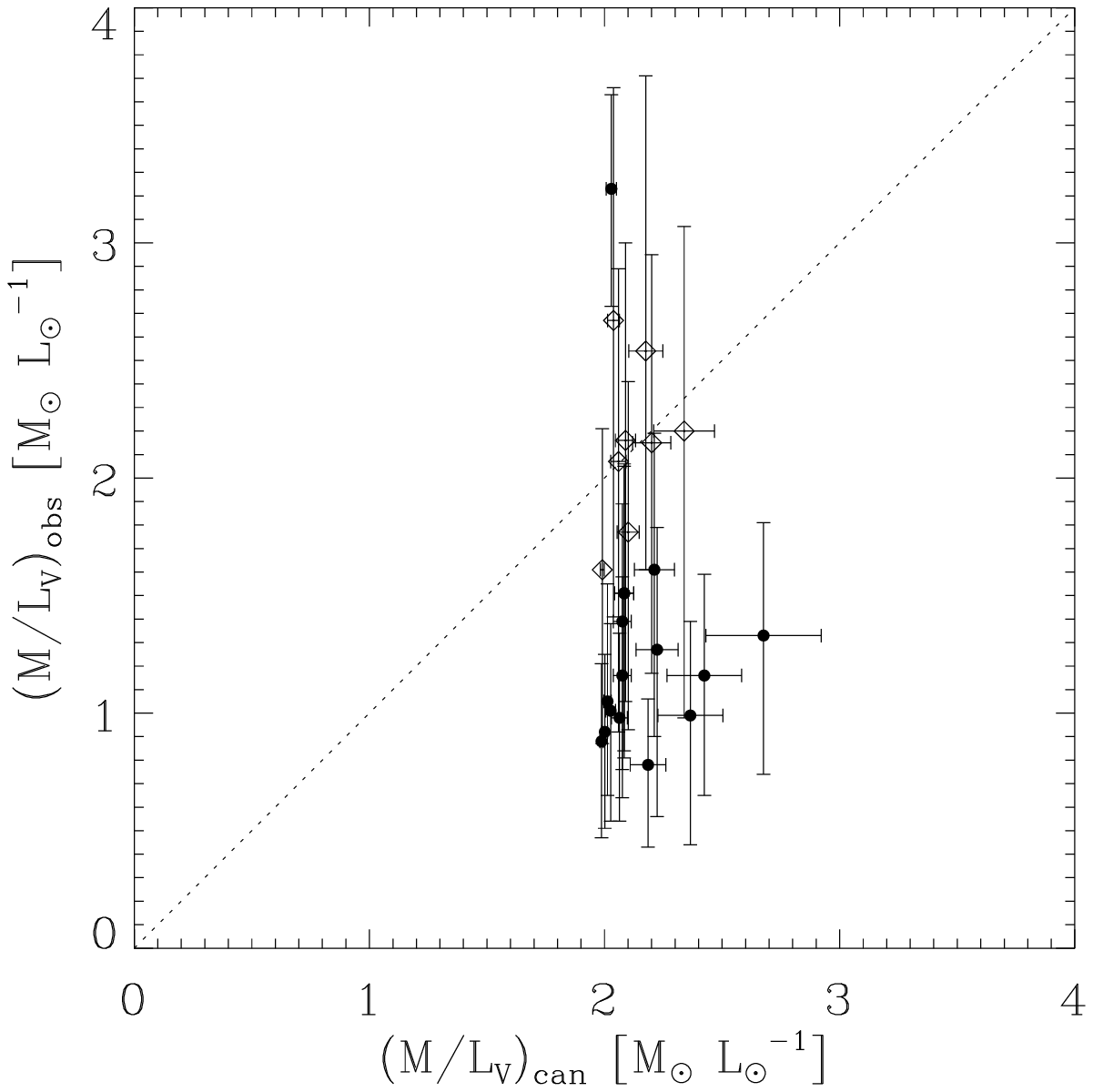}\includegraphics{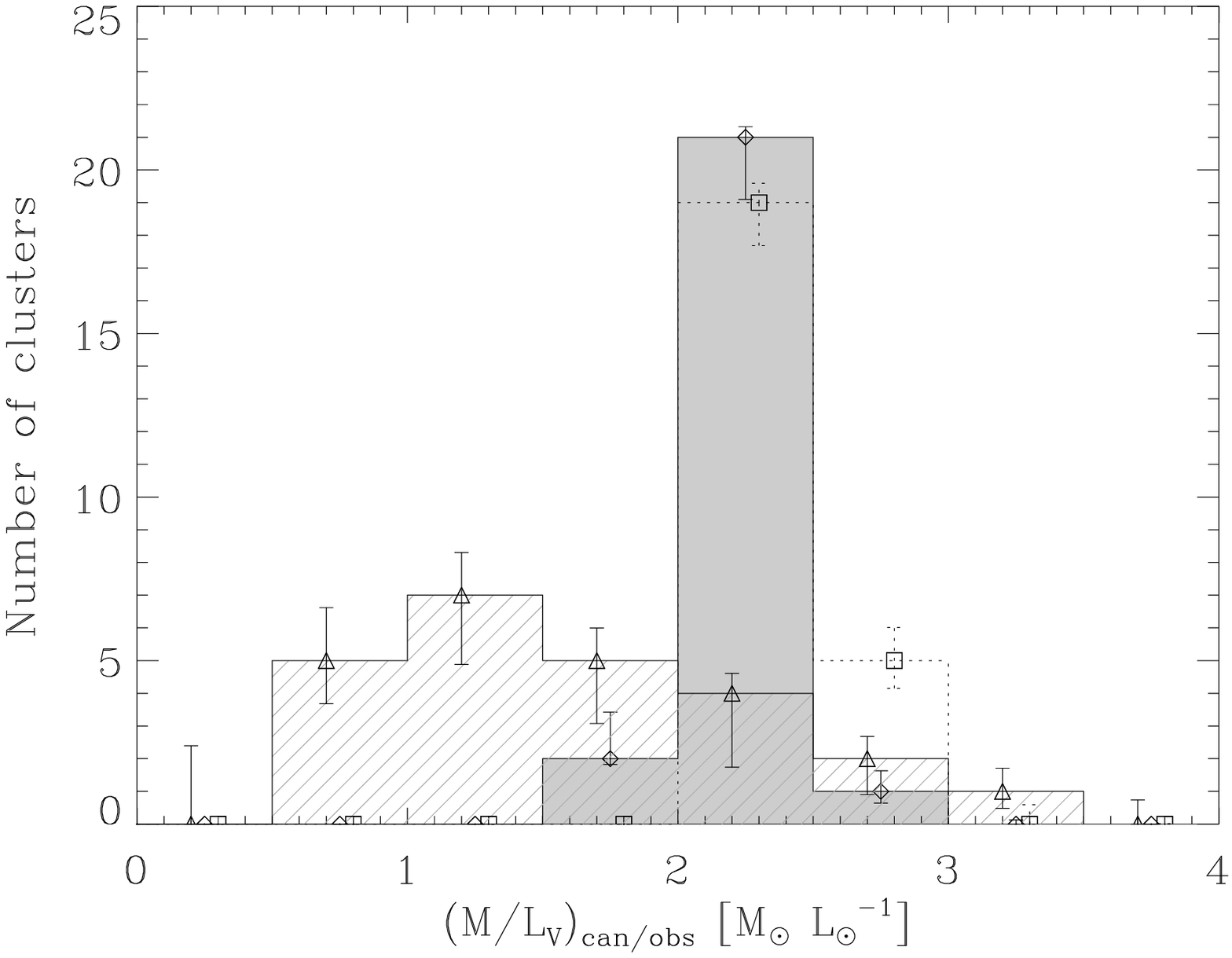}}
\caption[]{\label{fig:obs}
       {\it Left}: Observed mass-to-light ratio $(M/L_V)_{\rm obs}$ versus the canonically expected mass-to-light ratio $(M/L_V)_{\rm can}$, together with their $1\sigma$ standard errors (see Table~\ref{tab:obs}). The dotted line follows the 1:1 relation. Clusters for which the disagreement is larger than 1$\sigma$ are plotted as dots. {\it Right}: Number histogram of $(M/L_V)_{\rm can}$ (diamonds, shaded area) and $(M/L_V)_{\rm obs}$ (triangles, hashed area). For comparison, canonically expected $M/L_V$ ratios from \citet{bruzual03} for a \citet{chabrier03} IMF are overplotted (squares, dotted line). Again, the error bars denote $1\sigma$ deviations, which were determined from 30k random realisations of the underlying data.
          }
\end{figure*}

In Table~\ref{tab:obs} the observed properties are listed of the 24 Galactic globular clusters for which the $V$-band mass-to-light ratios ($M/L_V$) and orbital parameters are available, i.e., the sample that is covered both by \citet{dinescu99} and \citet{mclaughlin05}. A first inspection of the observed $M/L_V$ ratios can be made by comparing them to the (`canonical') $M/L$ ratios from SSP models, which only depend on metallicity due to the invariance of the shape of the stellar mass function in these models. In Fig.~\ref{fig:obs} (left), the observed $M/L$ ratios of our sample are plotted versus the canonically expected values that were computed by interpolating SSP models. These were emulated with the models from KL08 neglecting the preferential loss of low-mass stars. The discrepancy between observed and expected $M/L_V$ ratio is evident, as the canonical $M/L_V$ are constrained to a much narrower and generally higher range than the observed ones. The number histogram of the two $M/L_V$ (Fig.~\ref{fig:obs}, right) further substantiates this dissimilarity. The observed $M/L_V$ ratios are on average 74$^{+6}_{-7}$\% of the canonically expected values.

With the present paper we aim to quantify the contribution of dynamical effects such as the preferential loss of low-mass stars and the selective loss of stellar remnants (see KL08) to the discrepancy between the observed and canonically expected $M/L$ ratios. In Sect.~\ref{sec:clevo}, we summarise the cluster models from KL08 and highlight the aspects that are particularly relevant to this study. The dissolution timescales for the cluster sample are computed in Sect.~\ref{sec:t0}, whereas the predicted mass and $M/L_V$ evolution are considered and compared to the observations in Sect.~\ref{sec:pred}. We predict slopes of the low-mass stellar mass function and discuss observational tests to verify the preferential loss hypothesis for appropriate clusters in Sect.~\ref{sec:test}. In the final Sect.~\ref{sec:disc}, we discuss the results and present our conclusions.

\section{Cluster evolution models and $M/L_V$ evolution} \label{sec:clevo}
In order to study the evolution of clusters on specific orbits, we use analytical cluster models ({\tt SPACE}, see KL08) that incorporate the effects of stellar evolution, stellar remnant production, cluster dissolution and energy equipartition. They are summarised here and are treated in more detail in KL08. In the second part of this section, the dependence of mass-to-light ratio evolution on initial mass, metallicity and dissolution timescale is assessed \citep[for a more detailed description, see][]{kruijssen08b}.

\subsection{Summary of the models} \label{sec:model}
In the {\tt SPACE} cluster models, clusters gradually lose mass due to stellar evolution and dissolution. The total cluster mass evolution is determined by 
\begin{equation}
\label{eq:dmdt}
\frac{\d\mc}{\d t}=\left(\frac{\d\mc}{\d t}\right)_{\rm ev}+\left(\frac{\d\mc}{\d t}\right)_{\rm dis} ,
\end{equation}
where the first term denotes mass loss due to stellar evolution and the second represents mass loss by dissolution. Additionally, the formation of stellar remnants and the mass-dependent loss of stars by dissolution are taken into account, thus providing a description of the changing mass function and cluster mass in remnants.

Stellar evolution is included by using the Padova 1999 isochrones\footnote{These isochrones are based on \citet{bertelli94}, but use the AGB treatment as in \citet{girardi00}.}. These are available for metallicities $Z=\{0.0004,0.004,0.008,0.02,0.05\}$ (or [Fe/H]$=\{-1.7,-0.7,-0.4,0.0,0.4\}$), which thus restricts our model computations to these values. Stellar evolution removes the most massive stars from the cluster and increases the non-luminous cluster mass by turning stars into remnants, which is included by adopting an initial-remnant mass relation\footnote{For white dwarfs, this relation is taken from \citet{kalirai07}, while for neutron stars the relation from \citet{nomoto88} is used. Black hole masses are assumed to be constant at $8~\msun$, in agreement with observations \citep{casares07}. For more details, see KL08.}. A \citet{kroupa01} IMF is assumed.

Cluster dissolution represents the dynamical cluster mass loss due to stars passing the tidal radius, which acts on the timescale $\tau_{\rm dis}$:
\begin{equation}
\label{eq:dmdtdis}
  \left(\frac{\d \mct}{\d t}\right)_{\rm dis} = -\frac{\mct}{\tau_{\rm dis}} = -\frac{\mct^{1-\gamma}}{t_0} ,
\end{equation}
where $\mct$ represents the present day cluster mass and the second equality follows from the relation derived by \citet{lamers05}:
\begin{equation}
\label{eq:tdis}
\tau_{\rm dis}=t_0[\mct/\msun]^\gamma .
\end{equation}
The characteristic timescale $t_0$ depends on the environment and determines the strength of dissolution. For example, {in the case of dissolution by two-body relaxation}, $t_0$ depends on tidal field strength and therefore on the angular velocity of the cluster orbit. Typical values are $t_0=10^5$---$10^8$~yr \citep[e.g.,][]{lamers05a}, translating into a total disruption time $\tdis\approx 10^8$---$10^{11}$~yr for a 10$^5~\msun$ cluster. The exponent $\gamma$ is found to be $\gamma\approx 0.62$, both from observations \citep{boutloukos03,gieles05a} and from the \citet{baumgardt03} $N$-body simulations of tidal dissolution {for clusters with King parameter $W_0=5$} \citep{lamers05a}. {However, it is recently derived by \citet{lamers09} that $\gamma=0.70$ for King parameter $W_0=7$. Since this concentration more closely resembles the mean King parameter for Galactic GCs (see Table~\ref{tab:obs}), we adopt $\gamma=0.70$ throughout this study.}

The effect of dissolution on the mass function depends on the dynamical state of the cluster. {As it evolves towards} energy equipartition, low-mass stars are preferentially lost from the cluster. This mass loss (in the `preferential mode', KL08) is approximated by increasing the minimum stellar mass \citep{lamers06}, while evaporation that is independent of stellar mass (mass loss in the `canonical mode', KL08) is accounted for by decreasing the normalisation of the mass function. In our models, both modes coexist to allow for intermediate modes of mass loss. {Their relative contributions are fitted such that the $M/L_V$ ratio evolution matches the $N$-body simulations by \citet{baumgardt03}.}

Cluster photometry is computed by integrating the stellar mass function over the {stellar} isochrones, yielding cluster magnitude evolution $M_{\rm \lambda}(t,\mci)$ for a passband $\lambda$ and a cluster with initial mass $\mci$.

\subsection{Dependence of mass-to-light ratio on model parameters} \label{sec:depend}
\begin{figure}[t]
\resizebox{\hsize}{!}{\includegraphics{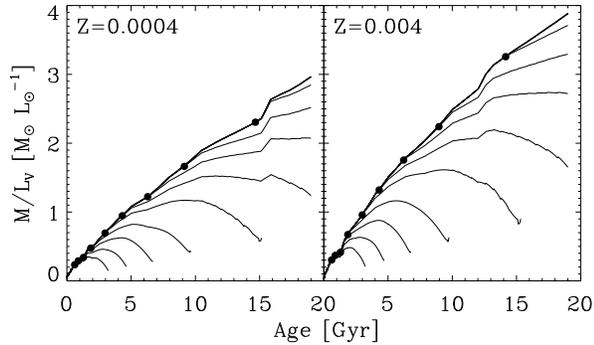}}
\caption[]{\label{fig:model1}
       {\it Left}: $V$-band mass-to-light ratio evolution $M/L_V$ for $t_0=1$~Myr, $Z=0.0004$ and initial masses in the range $\log{\mcli}=5$---8 with 0.25-dex intervals. {\it Right}: same graph, but for $Z=0.004$. From top to bottom, different curves represent the $M/L_V$ evolution for decreasing initial cluster masses.
          }
\end{figure}

The models described in Sect.~\ref{sec:model} yield a mass-to-light ratio evolution that depends on the dissolution timescale, metallicity and initial cluster mass. In canonical models, i.e., without the preferential loss of low-mass stars, $M/L$ monotonously increases with time. For a given age and metallicity, these models provide $M/L$ ratios that are independent of cluster mass. On the other hand, our models including dynamical effects predict a mass-dependent drop in mass-to-light ratio due to the ejection of low-mass, high-$M/L$ stars \citep[KL08]{kruijssen08b}. In Fig.~\ref{fig:model1}, the $V$-band mass-to-light ratio evolution $M/L_V$ is shown for two metallicities and several initial cluster masses. In both panels, the upper curve marks the canonical mass-to-light ratio evolution, while the others correspond to cluster evolution including the preferential loss of low-mass stars for different initial masses. Since low-mass clusters evolve on shorter timescales than massive ones, the deviation of their mass-to-light ratio evolution with respect to canonical models arises at earlier times than for massive clusters.

{The mass-to-light ratio decrease can be quantified by considering the ratio of the observed or predicted $M/L$ to its canonical value to divide out their metallicity dependence:
\begin{equation}
  \label{eq:q}
  \mathcal{Q}_{\rm obs/pred}\equiv\frac{(M/L_V)_{\rm obs/pred}}{(M/L_V)_{\rm can}} .
\end{equation}
Figure~\ref{fig:mldroptdis} shows the predicted fraction of the canonical $M/L$ ratio $\mathcal{Q}_{\rm pred}$ as a function of $t/\tdis$ for clusters with initial masses in the range $\mcli=2\times 10^4$---$10^8~\msun$, dissolution timescales $t_0=\{1,10\}$~Myr and metallicities $Z=\{0.0004,0.004\}$. It shows that $\mathcal{Q}_{\rm pred}$ is independent of metallicity, initial cluster mass and dissolution timescale when considered as a function of the elapsed fraction of the total disruption time $t/\tdis$. The three-component linear approximation illustrates the well-defined uniform correlation and is given by
\begin{equation}
  \label{eq:qtdis}
   \mathcal{Q}_{\rm pred} = \left\{
  \begin{array}{ccl}
  1 & {\rm for} & t/\tdis < 0.2,  \\
  1.142-0.71t/\tdis & {\rm for} & 0.2\leq t/\tdis < 0.7,  \\
  1.471-1.18t/\tdis & {\rm for} & t/\tdis \geq 0.7, \\
  \end{array}\right.
\end{equation}
which applies for all initial conditions, i.e., is independent of the cluster properties or environment\footnote{Please note that a Kroupa IMF was assumed here. For substantially different IMFs the relation will vary.}. Equation~\ref{eq:qtdis} does not include possible effects of primordial mass segregation on the change of the mass function. From model runs where we did assume the preferential depletion of low-mass stars from $t=0$ on we know that its effects become about 10\% stronger with respect to purely dynamically induced low-mass star depletion (KL08). This number should be treated with some care, because our models are based on $N$-body simulations of clusters that did not start out in a mass-segregated state \citep{baumgardt03}.
\begin{figure}[t]
\resizebox{\hsize}{!}{\includegraphics{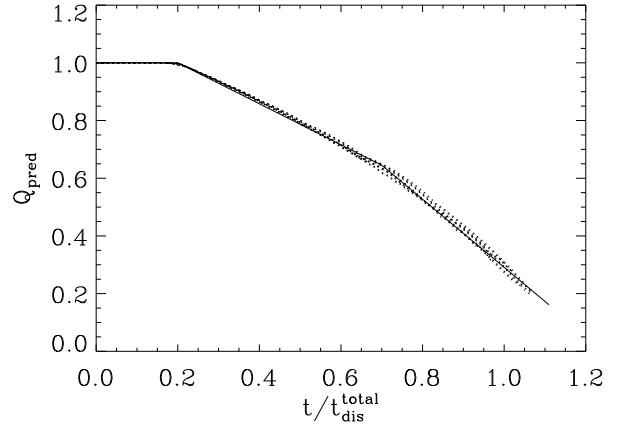}}
\caption[]{\label{fig:mldroptdis}
      The ratio of predicted to {canonical} mass-to-light ratio $\mathcal{Q}_{\rm pred}$ as a function of the {elapsed} fraction of the total disruption time $t/\tdis$. Dotted curves denote model predictions for a broad range of initial conditions (varying initial masses, dissolution timescales and metallicities), while the solid line describes a three-component linear approximation to the models.
          }
\end{figure}

\begin{figure}[t]
\resizebox{\hsize}{!}{\includegraphics{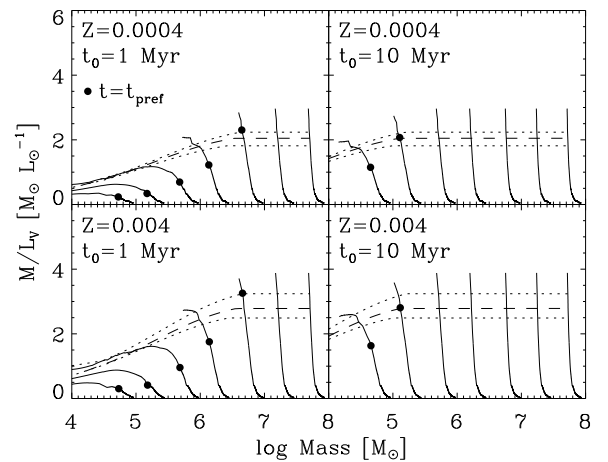}}
\caption[]{\label{fig:model2}
      Cluster evolution in the $\{M,M/L_V\}$-plane for $t_0=\{1,10\}$~Myr and $Z=\{0.0004,0.004\}$. Solid curves represent cluster evolutionary tracks for initial cluster masses in the range $\mcli=10^5$---$10^8$~\msun with 0.5-dex intervals. Cluster isochrones at $t=12$~Gyr are described by dashed lines, while these at $t=10$~and $t=14$~Gyr are denoted by dotted lines (bottom and top, respectively). Dots denote the onset of the preferential loss of low-mass stars for each evolutionary track.
          }
\end{figure}
The relation between the predicted fraction of the canonical mass-to-light ratio $\mathcal{Q_{\rm pred}}$ and the elapsed fraction of the total disruption time is expected, since in our models the decrease of $M/L$ is the result of dynamical evolution. It is in agreement with studies by \citet{richer91} and \citet{baumgardt03}, who find that the depletion of the low-mass stellar mass function in globular clusters is closely related to the elapsed fraction of the total disruption time. Considering the physical processes driving dissolution, the result is not surprising either. Two-body relaxation is known to preferentially eject low-mass stars \citep{henon69} and tidal shocks remove the outer parts of the cluster, which in the case of mass segregation are constituted by low-mass stars.}

\begin{table*}[tb]\centering
\begin{tabular}{c |c c c c c c}
  \hline \multicolumn{7}{c}{Orbital parameters} \\
  \hline \hline NGC & $R_{\rm a}^\star$ (kpc) & $R_{\rm p}^\star$ (kpc) & $e^\star$ & P$^\star$ (Myr) & $V_{\rm c,a}^\dagger$ (km s$^{-1}$) & $\nu_{\rm sh}^\star$ (10 Gyr)$^{-1}$  \\ \hline
    104 & 7.3 $\pm$   0.1    &   5.3 $\pm$ 0.3 &  0.16 $\pm$ 0.04  & 193  $\pm$ 4   &    221.4 $\pm$ 0.2 &      $(0.501\pm 0.134)\times 10^{-2}$          \\
    288 & 11.1 $\pm$  0.4    &   1.8 $\pm$ 0.5 &  0.72 $\pm$ 0.06  & 237  $\pm$ 12  &    213.5 $\pm$ 0.8 &    $(0.739\pm 0.186)\times 10^{0}$           \\
  1851 & 34.7 $\pm$ 5.9     &   5.7 $\pm$ 1.2 &  0.72 $\pm$ 0.02 & 685  $\pm$ 114 &    195.9 $\pm$ 1.4 &    $(0.804\pm 0.286)\times 10^{-3}$             \\
  1904 & 20.4 $\pm$ 1.3     &   4.4 $\pm$ 1.7 &  0.64 $\pm$ 0.10 & 422  $\pm$ 32  &    201.9 $\pm$ 1.0 &     $(0.592\pm 0.540)\times 10^{-2}$            \\
  4147 & 26.8 $\pm$ 3.4     &   4.0 $\pm$ 1.9 &  0.74 $\pm$ 0.08 & 551  $\pm$ 74  &    198.4 $\pm$ 1.4 &     $(0.208\pm 0.098)\times 10^{-1}$            \\
  4590 & 30.0 $\pm$ 3.7     &   8.7 $\pm$ 0.4 &  0.55 $\pm$ 0.03 & 650  $\pm$ 78  &    197.2 $\pm$ 1.2 &     $(0.208\pm 0.050)\times 10^{-2}$           \\
  5139 & 6.4  $\pm$    0.1   &   1.2 $\pm$ 0.1 &  0.69 $\pm$  0.02 & 123  $\pm$ 1   &    222.8 $\pm$ 0.1 &      $(0.373\pm 0.082)\times 10^{0}$         \\
  5272 & 14.0 $\pm$ 0.8     &   5.4 $\pm$ 0.8 &  0.44 $\pm$ 0.06 & 321  $\pm$ 18  &    208.6 $\pm$ 1.2 &     $(0.209\pm 0.108)\times 10^{-2}$            \\
  5466 & 69.8 $\pm$ 29.6   &   6.7 $\pm$ 1.4 &  0.83 $\pm$ 0.03 & 1340 $\pm$ 595 &    192.1 $\pm$ 1.4&   $(0.110\pm 0.074)\times 10^{0}$         \\
  5904 & 46.1 $\pm$ 12.8   &   2.5 $\pm$ 0.2 &  0.90 $\pm$ 0.02 & 995  $\pm$ 286 &    193.9 $\pm$ 1.6 &   $(0.117\pm 0.048)\times 10^{-1}$         \\
  6093 & 3.2  $\pm$   0.2    &   1.0 $\pm$ 0.6 &  0.54 $\pm$ 0.21  & 65   $\pm$ 6   &    213.3 $\pm$ 1.8 &       $(0.121\pm 0.085)\times 10^{-1}$        \\
  6121 & 5.8  $\pm$    0.3   &   0.7 $\pm$ 0.1 &  0.79 $\pm$ 0.03  & 114  $\pm$ 3   &    223.2 $\pm$ 0.1 &      $(0.280\pm 0.072)\times 10^{0}$         \\
  6171 & 3.3  $\pm$   0.2    &   2.8 $\pm$ 0.3 &  0.08 $\pm$ 0.07  & 99   $\pm$ 7   &    214.2 $\pm$ 1.7 &       $(0.125\pm 0.037)\times 10^{0}$        \\
  6205 & 25.3 $\pm$  6.9    &   5.7 $\pm$ 0.5 & 0.63 $\pm$ 0.07  & 526  $\pm$ 132 &    199.0 $\pm$ 3.3  &   $(0.154\pm 0.068)\times 10^{-2}$            \\
  6218 & 5.3  $\pm$     0.1  &   2.8 $\pm$ 0.3 &  0.30 $\pm$ 0.04  & 130  $\pm$ 4   &    223.0 $\pm$ 0.1 &      $(0.235\pm 0.075)\times 10^{-1}$          \\
  6254 & 5.0  $\pm$     0.2  &   3.4 $\pm$ 0.4 &  0.18 $\pm$ 0.05  & 132  $\pm$ 7   &    222.7 $\pm$ 0.3 &      $(0.261\pm 0.074)\times 10^{-1}$         \\
  6341 & 9.9  $\pm$   0.4    &   1.3 $\pm$ 0.1 &  0.78 $\pm$ 0.03  & 208  $\pm$ 12  &    216.0 $\pm$ 0.8 &    $(0.167\pm 0.060)\times 10^{-1}$          \\
  6362 & 5.3  $\pm$     0.1  &   2.6 $\pm$ 0.2 &  0.35 $\pm$ 0.04  & 124  $\pm$ 2   &    223.0 $\pm$ 0.1 &      $(0.491\pm 0.125)\times 10^{0}$         \\
  6656 & 9.6  $\pm$   0.7    &   2.8 $\pm$ 0.2 &  0.55 $\pm$ 0.01  & 197  $\pm$ 14  &    216.6 $\pm$ 1.5 &    $(0.441\pm 0.114)\times 10^{-1}$           \\
  6712 & 5.9  $\pm$   0.3    &   0.9 $\pm$ 0.1 &  0.74 $\pm$ 0.04  & 126  $\pm$ 11  &    223.2 $\pm$ 0.1 &    $(0.114\pm 0.041)\times 10^{0}$        \\
  6779 & 13.0 $\pm$  1.9    &   0.8 $\pm$ 0.3 &  0.88 $\pm$ 0.03 & 249  $\pm$ 30  &    210.2 $\pm$ 3.1  &    $(0.407\pm 0.142)\times 10^{0}$          \\
  6809 & 6.0  $\pm$     0.3   &   1.7 $\pm$ 0.2 &  0.56 $\pm$ 0.04 & 136  $\pm$ 5   &    223.1 $\pm$ 0.2 &      $(0.177\pm 0.048)\times 10^{0}$         \\
  6934 & 46.8 $\pm$  19.8  &   6.7 $\pm$ 1.6 &  0.75 $\pm$ 0.06 & 990  $\pm$ 434 &    193.8 $\pm$ 2.4 &   $(0.149\pm 0.112)\times 10^{-2}$          \\
  7089 & 42.2 $\pm$  17.9  &   6.3 $\pm$ 1.2 &  0.74 $\pm$ 0.06 & 860  $\pm$ 379 &    194.5 $\pm$ 2.7 &   $(0.818\pm1.220)\times 10^{-3}$          \\\hline
\end{tabular}
\caption[]{\label{tab:orbits}
    Orbital parameters for the cluster sample, together with their $1\sigma$ standard errors. Consecutive columns list the cluster NGC number, apogalactic radius $R_{\rm a}$, perigalactic radius $R_{\rm p}$, eccentricity $e$, orbital period $P$, circular velocity of the gravitational potential at the distance of apogalacticon $V_{\rm c,a}$ and destruction rate due to disc shocking $\nu_{\rm sh}$. The circular velocities are computed in the galactic plane ($z=0$). Because the gravitational potentials of the disc and bulge decrease with $|z|$, this implies that for clusters with $R_{\rm a}<10$~kpc the actual $V_{\rm c,a}$ can be 5---15\% lower.
     \\ $^\star$From \citet{dinescu99}.
     \\ $^\dagger$Computed using the galactic potential from \citet{paczynski90}.
    }
\end{table*}
The evolution of cluster mass and $M/L_V$ can both be considered in the $\{M,M/L_V\}$-plane. The resulting `evolutionary tracks' are shown in Fig.~\ref{fig:model2} for two different dissolution timescales and again for two metallicities and a range of initial cluster masses as in Fig.~\ref{fig:model1}. Clusters start with their initial masses and with $M/L_V$ ratios close to zero, corresponding to an initial position on the $x$-axis of Fig.~\ref{fig:model2}. As time progresses, clusters initially evolve to lower masses and increasing $M/L_V$ due to the death of massive stars, translating into up- and leftward motion in the $\{M,M/L_V\}$-plane. When the preferential loss of low-mass stars becomes an important mechanism (the onset of which is marked by dots for each evolutionary track), the $M/L_V$ increase is turned into a decrease instead, as also illustrated in Fig.~\ref{fig:model1}. In Fig.~\ref{fig:model2}, the thus attained maximum in the $M/L_V$ evolution is best visible for low cluster masses and $t_0=1$~Myr.

Since Galactic globular clusters generally share the same ages \citep[e.g.,][]{vandenberg90}, the observed distribution of GCs in the $\{M,M/L_V\}$-plane would follow curves of equal age in Fig.~\ref{fig:model2} if there were no spreads in metallicity and dissolution timescale. These curves, or cluster isochrones, are shown for ages $t=\{10,12,14\}$~Gyr. Along the isochrones, $M/L_V$ increases with cluster mass since massive clusters have spent {a smaller fraction of their total disruption time} than low-mass clusters and will therefore have experienced a smaller $M/L_V$ decrease due to low-mass star depletion. The curves flatten at the highest masses, since these clusters have not yet exhibited significant preferential low-mass star ejection.

From Fig.~\ref{fig:model2} we infer the influences of dissolution timescale and metallicity on the mass-to-light ratio evolution. The dissolution timescale sets the cluster mass for which the down-bend of the cluster evolutionary tracks can occur and therefore also determines the location of the `knee' in the cluster isochrones. The metallicity determines the vertical extent of the cluster evolutionary tracks and thus the $M/L_V$-normalisation of the cluster isochrones. {As set forth in \citet{kruijssen08b}, the natural spread in dissolution timescale and metallicity thus explains the scatter around the relation between $M/L$ and cluster mass observed by \citet{mandushev91}.}

\section{Determining the dissolution timescale} \label{sec:t0}
To assess the influence of the preferential loss of low-mass stars on the low observed mass-to-light ratios, the orbital parameters of individual clusters are to be translated into the appropriate dissolution timescales $t_0$ for use in our cluster models. The computation is treated in this section.

\subsection{Cluster dissolution timescales from orbital parameters}
For globular clusters, {dissolution due to two-body relaxation in the Galactic tidal field\footnote{This includes the effect of bulge shocks, which occur in clusters on eccentric orbits.}} and disc shocking are the main dissolution mechanisms \citep[e.g.,][]{chernoff86}. The total dissolution timescale $t_{0,{\rm tot}}$ can be written as
\begin{equation}
  \label{eq:t0}
  \frac{1}{t_{0,{\rm tot}}}=\frac{1}{t_{0,{\rm evap}}}+\frac{1}{t_{0,{\rm sh}}} ,
\end{equation}
where $t_{0,{\rm evap}}$ denotes the dissolution timescale due to {two-body relaxation or evaporation (carrying the subscript `evap')} and $t_{0,{\rm sh}}$ the dissolution timescale due to disc shocking.

For the {dissolution timescale due to two-body relaxation}, we use the expression for the total disruption time from \citet[Eq. 10]{baumgardt03} as approximated by \citet{lamers05a} to write
\begin{equation}
  \label{eq:t0evap}
  t_{0,{\rm evap}}=t_{0,{\rm evap}}^\odot~\left(\frac{R_{\rm gc,a}}{8.5~{\rm kpc}}\right)\left(\frac{V_{\rm c,a}}{220~{\rm km~s^{-1}}}\right)^{-1}(1-e) ,
\end{equation}
with $t_{0,{\rm evap}}^\odot$ the dissolution timescale due to {two-body relaxation} for a circular orbit at the solar galactocentric radius, $R_{\rm gc,a}$ the apogalactic radius of the cluster orbit, $V_{\rm c,a}$ the circular velocity {of the gravitational potential at the distance of} apogalacticon and $e$ the orbital eccentricity. Values for $R_{\rm gc,a}$ are taken from \citet{dinescu99}, while the circular velocities are computed for the galactic potential from \citet{paczynski90}. This potential, as well as the one from \citet{johnston95}, is used by \citet{dinescu99} in the determination of the cluster orbits. By comparing our models to the $N$-body simulations by \citet{baumgardt03} we find $t_{0,{\rm evap}}^\odot=21.3$~Myr for clusters with $W_0=5$ King profiles, in very close agreement with earlier reported values of 20.9~Myr \citep{lamers05} and 22.8~Myr \citep{lamers06a}. {Using the same method for $\gamma=0.7$, corresponding to $W_0=7$ King profiles (see Sect.~\ref{sec:model}), we obtain $t_{0,{\rm evap}}^\odot=10.7$~Myr. This is the adopted value in this paper.}

The dissolution timescale due to disc shocking can be obtained from the globular cluster destruction rates due to disc shocking $\nu_{\rm sh}$ from \citet{dinescu99}. Following from Eq.~\ref{eq:tdis}, a present destruction rate $\nu(t)$ is related to a dissolution timescale $t_0$ by
\begin{equation}
  \label{eq:nu}
  \nu(t)=\frac{10^{10}}{t_0(\mct/\msun)^\gamma} ,
\end{equation}
with $\nu$ in units of $(10~{\rm Gyr})^{-1}$, $t_0$ in years, and $\mct$ denoting the cluster mass at age $t$. The denominator represents {an estimate for} the total cluster lifetime. This expression can be inverted to obtain $t_{0,{\rm sh}}$ from $\nu_{\rm sh}$. However, in \citet{dinescu99} constant $M/L_V=3~\msun~{\rm L}_\odot^{-1}$ is assumed to compute the cluster masses. Since their destruction rates are derived from a relation $\nu_{\rm sh}\propto M^{-1}$, these should be corrected for the actual mass-to-light ratios. We define the correction factor
\begin{equation}
  \label{eq:mlcorr}
  x_{\rm corr}=\frac{(M/L_V)_{\rm cst}}{(M/L_V)_{\rm obs}} ,
\end{equation}
with the numerator the constant mass-to-light ratio $(M/L_V)_{\rm cst}=3~\msun~{\rm L}_\odot^{-1}$ and the denominator the observed dynamical mass-to-light ratio from \citet{mclaughlin05} (see Table~\ref{tab:obs}). This allows us to express the dissolution timescale due to disc shocking as
\begin{equation}
  \label{eq:t0sh}
  t_{0,{\rm sh}}=\frac{10^{10}}{x_{\rm corr}\nu_{\rm sh}(\mct/\msun)^\gamma} .
\end{equation}
Substitution of Eqs.~\ref{eq:t0evap} and~\ref{eq:t0sh} into Eq.~\ref{eq:t0} then yields the total dissolution timescale $t_{0,{\rm tot}}$.

\begin{table}[tb]\centering
\begin{tabular}{c |c c c}
  \hline \multicolumn{4}{c}{Dissolution timescales ($\gamma=0.70$)} \\
  \hline \hline NGC & $t_{0,{\rm evap}}$ (Myr) & $t_{0,{\rm sh}}$ (Myr) & $t_{0,{\rm tot}}$ (Myr) \\ \hline
    104   &      $7.7 \pm 0.4 $               &      76.6$_{- 35.1 }^{+33.9}$           &          7.0$_{-0.4}^{+0.4}$                      \\
    288   &      $4.0 \pm 0.9 $           &          3.6$_{- 2.6  }^{+5.0}$            &           1.9$_{-0.7}^{+1.4}$                \\
  1851   &      $13.7 \pm 2.6$                    &   1095.3$_{- 545.0}^{+553.9}$         &        13.6$_{-2.6}^{+2.6}$                        \\
  1904   &      $10.1 \pm 2.9$                    &   211.9$_{- 136.4}^{+204.8}$           &        9.6$_{-2.6}^{+2.7}$                         \\
  4147   &      $9.7 \pm 3.3 $                  &    135.9$_{- 73.8 }^{+79.7}$            &        9.1$_{-2.9}^{+2.9}$                         \\
  4590   &      $19.0 \pm 2.8$                    &   827.6$_{- 375.1}^{+356.2}$            &       18.5$_{-2.6}^{+2.6}$                        \\
  5139   &      $2.5 \pm 0.2 $               &       0.6$_{- 0.2  }^{+0.3}$                &       0.5$_{-0.2}^{+0.2}$                     \\
  5272   &      $10.4 \pm 1.3$                    &   343.3$_{- 187.7}^{+208.5}$            &       10.1$_{-1.2}^{+1.2}$                         \\
  5466   &      $17.1 \pm 8.0$                    &   25.6$_{- 15.3 }^{+19.6}$               &      10.2$_{-3.8}^{+4.3}$                        \\
  5904   &      $6.6 \pm 2.3 $                  &    46.8$_{- 24.0 }^{+24.8}$              &       5.8$_{-1.8}^{+1.8}$                        \\
  6093   &      $1.9 \pm 0.9 $               &       88.9$_{- 88.4 }^{+12440.4}$           &      1.9$_{-0.8}^{+0.9}$                     \\
  6121   &      $1.5 \pm 0.2 $               &       6.0$_{- 3.8  }^{+5.1}$                &       1.2$_{-0.2}^{+0.3}$                       \\
  6171   &      $3.9 \pm 0.4 $               &       21.0$_{- 10.9 }^{+10.6}$               &      3.3$_{-0.4}^{+0.4}$                     \\
  6205   &      $13.0 \pm 4.5$                    &   485.3$_{- 257.4}^{+272.4}$             &      12.7$_{-4.3}^{+4.3}$                         \\
  6218   &      $4.6 \pm 0.3 $               &       90.6$_{- 44.3 }^{+42.5}$               &      4.4$_{-0.3}^{+0.3}$                       \\
  6254   &      $5.1 \pm 0.4 $                  &    59.8$_{- 29.1 }^{+28.0}$               &      4.7$_{-0.4}^{+0.4}$                     \\
  6341   &      $2.8 \pm 0.4 $               &       48.5$_{- 25.1 }^{+25.9}$               &      2.6$_{-0.4}^{+0.4}$                      \\
  6362   &      $4.3 \pm 0.3 $               &       3.6$_{- 1.6  }^{+1.6}$                 &      2.0$_{-0.5}^{+0.5}$                      \\
  6656   &      $5.5 \pm 0.5 $                  &    18.6$_{- 9.2  }^{+8.7}$                &      4.3$_{-0.6}^{+0.5}$                       \\
  6712   &      $1.9 \pm 0.3 $               &       10.7$_{- 9.5  }^{+52.9}$               &      1.6$_{-0.3}^{+0.6}$                    \\
  6779   &      $2.1 \pm 0.6 $               &       3.1$_{- 2.4  }^{+6.4}$                 &      1.2$_{-0.4}^{+1.0}$                      \\
  6809   &      $3.3 \pm 0.3 $               &       13.5$_{- 3.9  }^{+3.2}$                &      2.6$_{-0.3}^{+0.3}$                       \\
  6934   &      $16.7 \pm 8.3$                    &   910.7$_{- 556.0}^{+750.3}$               &    16.4$_{-8.0}^{+8.0}$                               \\
  7089   &      $15.6 \pm 7.7$                    &   511.3$_{- 371.0}^{+773.0}$               &    15.2$_{-7.3}^{+7.3}$                               \\\hline
\end{tabular}
\caption[]{\label{tab:t0}
     The dissolution timescales (for $\gamma=0.70$) due to {two-body relaxation} ($t_{0,{\rm evap}}$), disc shocking ($t_{0,{\rm sh}}$) and both mechanisms ($t_{0,{\rm tot}}$), together with their $1\sigma$ standard errors. All values are rounded to one decimal.
    }
\end{table}
In Table~\ref{tab:orbits}, our cluster sample is listed with the orbital parameters from \citet{dinescu99} for the \citet{paczynski90} potential and our computed circular velocities {of the gravitational potential at the distance of} apogalacticon. The corresponding dissolution timescales can be found in Table~\ref{tab:t0}. {The values for the dissolution timescale range from $t_{0,{\rm tot}}=0.5$---20~Myr, corresponding to total disruption times for a 10$^6~\msun$ cluster in the range $\tdis=8$---300~Gyr. This is is in good agreement with the range that is required for low-mass star depletion to explain the observed mass-to-light ratio drop \citep{kruijssen08b}.} By comparing the dissolution timescales for {two-body relaxation} $t_{0,{\rm evap}}$ and disc shocking $t_{0,{\rm sh}}$, {we can see that the latter destruction mechanism is important (i.e., lowers the total dissolution timescale $t_{0,{\rm tot}}$ by more than 40\% with respect to $t_{0,{\rm evap}}$) for the clusters NGC~288, 5139 ($\omega$Cen), 6362, and 6779.} These clusters all have perigalactic radii smaller than 3~kpc (see Table~\ref{tab:orbits}). For the error analysis of Tables~\ref{tab:orbits} and~\ref{tab:t0} and of the rest of this paper we refer to Appendix~\ref{sec:app}.

\section{Predicted and observed mass-to-light ratios} \label{sec:pred}
In this section we combine our cluster models and the derived dissolution timescales to study the mass-to-light ratio evolution for our sample of 24 Galactic globular clusters. Present-day $M/L_V$ ratios are predicted for the cluster sample and are compared to the observations. We also discuss the possible causes for the individual clusters that still lack convincing agreement.

\subsection{Predicted mass-to-light ratios for the cluster sample}
We employ the cluster models treated in Sects.~\ref{sec:clevo} and~\ref{sec:t0} to predict $M/L_V$ ratios for the cluster sample. The input parameters for the models are the dissolution timescale $t_0$ and metallicity $Z$. The latter is derived from the iron abundance [Fe/H] (see Table~\ref{tab:obs}) according to
\begin{equation}
  \label{eq:z}
  Z = Z_\odot\times 10^{\rm [Fe/H]} ,
\end{equation}
with $Z_\odot=0.02$, while the dissolution timescale is taken from Table~\ref{tab:t0} ($t_{0,{\rm tot}}$). Since all clusters in the sample have metallicities $Z<0.004$, for each cluster the models are computed with metallicities $Z=\{0.0004,0.004\}$ and the appropriate dissolution timescales. The evolution is computed for a grid of initial cluster masses, yielding cluster evolution tracks for the mass and $V$-band mass-to-light ratio $M/L_V$. For both metallicities, at $t=12$~Gyr the tracks are interpolated over the mass grid to match the observed cluster mass. This provides predictions for $M/L_V$, the initial cluster mass $\mcli$ and the total disruption time $\tdis$ for two metallicities. These are then interpolated over metallicity to obtain the model predictions for the appropriate metallicity.

\begin{figure}[t]
\resizebox{\hsize}{!}{\includegraphics{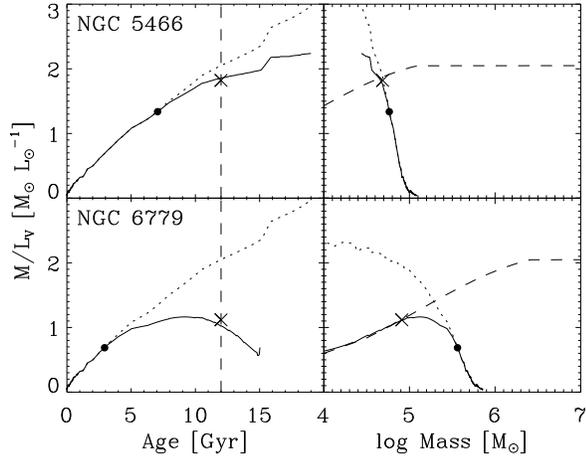}}
\caption[]{\label{fig:ngcs}
      {\it Top left}: Time-evolution of $M/L_V$ for NGC 5466. The solid line represents the $Z=0.0004$ model with the dissolution timescale of the cluster ($t_0=10.2$~Myr), while the dotted curve indicates the canonical $M/L_V$ evolution, i.e., if the preferential loss of low-mass stars were omitted. The dashed line denotes constant age of $t=12$~Gyr. The predicted $M/L_V$ of NGC~5466 is marked by a cross and the onset of the preferential loss of low-mass stars is specified with a dot. {\it Top right}: Evolution in the $\{M,M/L_V\}$-plane for NGC 5466. Curves and symbols have the same meaning as in the top-left panel. {\it Bottom left}: same as top left, but for NGC 6779 ($t_0=1.2$~Myr). {\it Bottom right}: same as top right, but for NGC 6779.
          }
\end{figure}
Examples of the $M/L_V$ evolution with time and mass are shown in Fig.~\ref{fig:ngcs} for NGC~5466 and~6779. The predicted $M/L_V$ are slightly offset with respect to the model curves because the models here are computed at $Z=0.0004$ while the predictions are interpolated over metallicity. However, the variation with metallicity is small for the displayed clusters, since their metallicities are close to $Z=0.0004$. It is evident that low-mass star depletion has a much stronger effect in the case of NGC~6779 than for NGC~5466. Considering their dissolution timescales ($t_0=1.2$~Myr versus $t_0=10.2$~Myr, respectively) and the resulting mass evolution, this is not surprising since NGC~6779 has suffered much stronger mass loss than NGC~5466.

\begin{figure}[t]
\resizebox{\hsize}{!}{\includegraphics{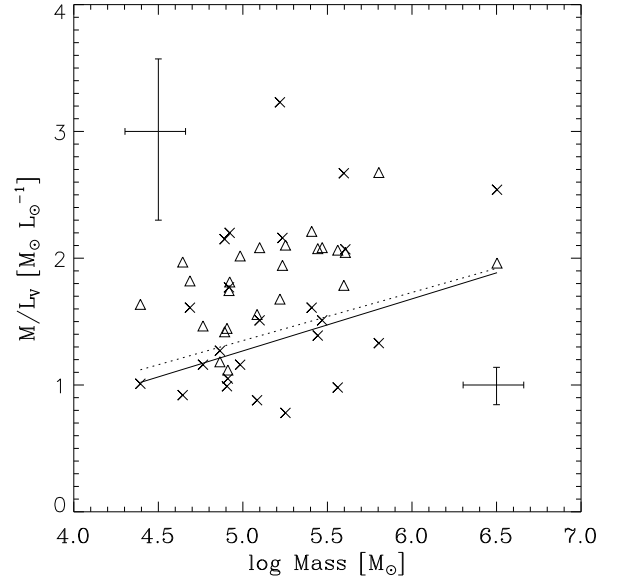}}
\caption[]{\label{fig:mmlplane}
      Observed (crosses) and predicted (triangles) distribution of GCs in the $\{M,M/L_V\}$-plane. {The solid line represents our linear fit to the observations, while the dotted line denotes the relation found by \citet{mandushev91}.}The error bars in the top left corner denote the average $1\sigma$ uncertainty on the observations, while the error bars in the bottom right corner represent the average $1\sigma$ uncertainty on the predictions.
          }
\end{figure}
The predicted mass-to-light ratios $M/L_V$, initial masses $\mcli$ and remaining lifetimes $\tdis-t$ are listed for our entire GC sample in Table~\ref{tab:pred}. In addition, the observed and predicted fractions of the canonical $M/L_V$ ratios $\mathcal{Q}_{\rm obs}$ and $\mathcal{Q}_{\rm pred}$ are shown, as well as the agreement between our predicted $M/L_V$ and the observed values. Combining Tables~\ref{tab:t0} and~\ref{tab:pred}, we see that GCs with short dissolution timescales indeed have low predicted $M/L$ ratios.
\begin{table*}[tb]\centering
\begin{tabular}{c |c c c c c c}
  \hline \multicolumn{7}{c}{Model predictions} \\
  \hline \hline NGC & $(M/L_V)_{\rm pred}$ (\msun~L$_\odot^{-1}$) & $\log{\mcli}$ (\msun) & $\tdis-t$ (Gyr)  & $\mathcal{Q}_{\rm obs}$ & $\mathcal{Q}_{\rm pred}$ & Agreement\\ \hline
    104     &  2.68$_{-0.25}^{+0.25}$    &  6.10$_{-0.18}^{+0.14}$    &  102.1$_{-32.4}^{+26.6}$    &  0.50$_{-0.23}^{+0.19}$   &  1.00$_{-0.00}^{+0.00}$   &   3     \\
    288     &  1.42$_{-0.29}^{+0.37}$    &  5.73$_{-0.26}^{+0.15}$    &  6.3$_{-2.7}^{+3.6}$             &  0.98$_{-0.45}^{+0.37}$   &  0.64$_{-0.13}^{+0.16}$   &   1    \\
  1851     & 2.21$_{-0.09}^{+0.09}$     &  5.71$_{-0.17}^{+0.14}$    &  105.4$_{-38.5}^{+33.5}$    & 0.73$_{-0.32}^{+0.26}$     &  1.00$_{-0.00}^{+0.00}$     &   2       \\
  1904     & 2.02$_{-0.13}^{+0.07}$     &  5.38$_{-0.16}^{+0.13}$    &  37.7$_{-16.8}^{+14.9}$      & 0.56$_{-0.25}^{+0.20}$     &  0.97$_{-0.06}^{+0.03}$     &   2       \\
  4147     & 1.64$_{-0.20}^{+0.16}$     &  4.99$_{-0.14}^{+0.12}$    &  13.2$_{-6.9}^{+6.1}$           &  0.50$_{-0.23}^{+0.18}$    &  0.81$_{-0.10}^{+0.08}$    &   2     \\
  4590     & 1.97$_{-0.09}^{+0.03}$     &  5.03$_{-0.16}^{+0.13}$    &  41.9$_{-15.9}^{+13.3}$      &  0.46$_{-0.20}^{+0.16}$     & 0.99$_{-0.05}^{+0.02}$   &   4 \\
  5139     & 1.96$_{-0.19}^{+0.19}$     &  6.98$_{-0.13}^{+0.15}$    &  22.9$_{-9.4}^{+11.9}$         &  1.17$_{-0.43}^{+0.54}$    &  0.90$_{-0.08}^{+0.08}$    &   1   \\
  5272     & 2.08$_{-0.04}^{+0.04}$     &  5.76$_{-0.18}^{+0.14}$    &  83.5$_{-29.1}^{+23.9}$      & 0.67$_{-0.30}^{+0.24}$     &  1.00$_{-0.00}^{+0.00}$     &   2     \\
  5466     & 1.82$_{-0.18}^{+0.13}$     &  5.15$_{-0.16}^{+0.13}$    &  24.7$_{-12.4}^{+12.3}$      & 0.81$_{-0.37}^{+0.30}$     &  0.91$_{-0.09}^{+0.07}$     &   1     \\
  5904     & 2.10$_{-0.16}^{+0.11}$     &  5.66$_{-0.16}^{+0.13}$    &  34.9$_{-15.7}^{+14.2}$      & 0.36$_{-0.16}^{+0.13}$     &  0.96$_{-0.07}^{+0.04}$     &   5    \\
  6093     & 1.79$_{-0.27}^{+0.19}$     &  6.10$_{-0.16}^{+0.14}$    &  19.5$_{-11.3}^{+11.3}$      &  1.31$_{-0.62}^{+0.49}$     & 0.88$_{-0.13}^{+0.09}$   &   1  \\
  6121     & 1.18$_{-0.22}^{+0.18}$     &  5.91$_{-0.11}^{+0.09}$    &  3.7$_{-1.8}^{+1.5}$             & 0.57$_{-0.32}^{+0.23}$     &  0.53$_{-0.09}^{+0.07}$     &   1    \\
  6171     & 1.81$_{-0.21}^{+0.17}$     &  5.56$_{-0.13}^{+0.10}$    &  11.3$_{-4.2}^{+3.2}$           &  0.94$_{-0.52}^{+0.38}$    &  0.78$_{-0.07}^{+0.05}$    &    1    \\
  6205     & 2.08$_{-0.04}^{+0.04}$     &  5.77$_{-0.18}^{+0.15}$    &  109.1$_{-49.9}^{+45.2}$    & 0.72$_{-0.34}^{+0.26}$     &  1.00$_{-0.00}^{+0.00}$     &   2     \\
  6218     & 1.74$_{-0.15}^{+0.12}$     &  5.48$_{-0.12}^{+0.10}$    &  15.0$_{-6.1}^{+4.7}$           &  0.84$_{-0.40}^{+0.31}$    &  0.83$_{-0.07}^{+0.05}$    &    1     \\
  6254     & 1.94$_{-0.12}^{+0.09}$     &  5.68$_{-0.16}^{+0.12}$    &  27.5$_{-12.0}^{+9.2}$         &  1.03$_{-0.53}^{+0.40}$    &  0.93$_{-0.05}^{+0.04}$    &    1     \\
  6341     & 1.56$_{-0.14}^{+0.12}$     &  5.71$_{-0.12}^{+0.10}$    &  11.8$_{-5.1}^{+4.2}$           &  0.44$_{-0.21}^{+0.17}$    &  0.78$_{-0.07}^{+0.06}$    &    2     \\
  6362     & 1.47$_{-0.24}^{+0.21}$     &  5.67$_{-0.11}^{+0.10}$    &  5.3$_{-2.2}^{+2.0}$             & 0.48$_{-0.21}^{+0.18}$     &  0.60$_{-0.07}^{+0.05}$     &   1    \\
  6656     & 2.05$_{-0.09}^{+0.03}$     &  5.98$_{-0.20}^{+0.14}$    &  45.9$_{-18.8}^{+14.0}$      &  1.01$_{-0.56}^{+0.40}$     & 0.99$_{-0.04}^{+0.01}$   &    1 \\
  6712     & 1.45$_{-0.25}^{+0.25}$     &  5.80$_{-0.16}^{+0.09}$    &  5.5$_{-2.4}^{+2.6}$             & 0.42$_{-0.23}^{+0.17}$     &  0.61$_{-0.09}^{+0.09}$     &   1    \\
  6779     & 1.12$_{-0.22}^{+0.38}$     &  5.91$_{-0.33}^{+0.15}$    &  4.1$_{-2.3}^{+4.2}$             & 0.52$_{-0.20}^{+0.25}$     &  0.56$_{-0.11}^{+0.19}$     &   1    \\
  6809     & 1.68$_{-0.07}^{+0.06}$     &  5.79$_{-0.05}^{+0.04}$    &  14.8$_{-2.3}^{+2.0}$           &  1.59$_{-0.25}^{+0.20}$    &  0.83$_{-0.03}^{+0.03}$    &    4    \\
  6934     & 2.08$_{-0.06}^{+0.04}$     &  5.42$_{-0.17}^{+0.14}$    &  77.7$_{-44.8}^{+42.5}$      & 0.72$_{-0.32}^{+0.26}$     &  1.00$_{-0.02}^{+0.00}$     &   2     \\
  7089     & 2.06$_{-0.03}^{+0.03}$     &  5.85$_{-0.18}^{+0.15}$    &  151.0$_{-85.3}^{+81.4}$    & 0.47$_{-0.21}^{+0.17}$     &  1.00$_{-0.00}^{+0.00}$     &   3     \\\hline
\end{tabular}
\caption[]{\label{tab:pred}
     Predicted $V$-band mass-to-light ratio $(M/L_V)_{\rm pred}$, logarithm of the initial cluster mass $\mcli$ and remaining lifetime $\tdis-t$ for the clusters under consideration, together with their $1\sigma$ standard errors. In the fifth and sixth column, respectively the ratios of observed to canonical mass-to-light ratio $\mathcal{Q}_{\rm obs}$  and predicted to canonical mass-to-light ratio $\mathcal{Q}_{\rm pred}$ are listed. The seventh column gives the level of agreement between the observed mass-to-light ratio $(M/L_V)_{\rm obs}$ and predicted mass-to-light ratio $(M/L_V)_{\rm pred}$, which is defined as follows. For $\Delta M/L_V\equiv (M/L_V)_{\rm pred}-(M/L_V)_{\rm obs}$, an agreement value of ``1'' means $|\Delta M/L_V|\leq 1\sigma$, ``2'' means $1\sigma <|\Delta M/L_V|\leq 2\sigma$, etc., with $\sigma^2\equiv\sigma_{(M/L)_{\rm obs}}^2+\sigma_{(M/L)_{\rm pred}}^2$.
    }
\end{table*}

\subsection{Comparison of predictions to observations} \label{sec:mlcomp}
The fifth column in Table~\ref{tab:pred} indicates the ratio between observed and predicted mass-to-light ratio $\mathcal{Q}_{\rm obs}\equiv (M/L_V)_{\rm obs}/(M/L_V)_{\rm can}$. Analogously, the sixth column gives the ratio between predicted and canonical mass-to-light ratio $\mathcal{Q}_{\rm pred}\equiv (M/L_V)_{\rm pred}/(M/L_V)_{\rm can}$. On average, the former ratio is 0.74$^{+0.06}_{-0.07}$, while the latter ratio is 0.85$\pm0.01$ for the 24 GCs investigated. {There are factors that introduce biases when comparing the predictions to the observations. Specifically, the observations are likely biased to central $M/L$ ratios for some GCs, while we predict global values. In Sect.~\ref{sec:unexpl} a more detailed consideration is provided in which the comparison of the predictions to the observations is refined.}

The seventh column in Table~\ref{tab:pred} gives the level of
agreement between the observed and predicted mass-to-light ratios,
which is defined as $n$ if $(n-1)\sigma <|\Delta M/L_V|\leq n\sigma$
for $\Delta M/L_V\equiv (M/L_V)_{\rm pred}-(M/L_V)_{\rm obs}$. Within
the $1\sigma$ uncertainty, the predicted $M/L_V$ agree with the
observed values for 12 clusters out of the 24-cluster
sample. {A Gaussian distribution of errors would yield an expected 16 out of 24 clusters to be found within $1\sigma$.}

As a first comparison and analogously to the presentation in \citet{mandushev91} and \citet{rejkuba07}, in Fig.~\ref{fig:mmlplane} the distribution of GCs in the $\{M,M/L_V\}$-plane is shown for the observed and predicted mass-to-light ratios. Both populations fall within the same range and follow comparable trends of increasing $M/L_V$ with cluster mass. {\citet{mandushev91} already provided an expression for the observed logarithm of the mass as a function of magnitude, which allows for a derivation of the expected trend in Fig.~\ref{fig:mmlplane}. They fit
\begin{equation}
  \label{eq:logm}
 \log{(M/\msun)}=(-0.456\pm0.024)M_V+(1.64\pm0.21) ,
\end{equation}
where $M_V$ represents the $V$-band absolute magnitude of the cluster. Adopting a solar value of $M_{V,\odot}=4.83$, the relation between $M/L_V$ and mass from \citet{mandushev91} can then be expressed as
\begin{equation}
  \label{eq:logml}
 \log{(M/L_V)}=(-0.12\pm0.05)\log{(M/\msun)}-(0.49\pm0.21) .
\end{equation}
A first-order Taylor expansion of $M/L_V$ around $\log{(M/\msun)}=5.2$ then gives 
\begin{equation}
  \label{eq:linml}
  M/L_V\approx0.38\log{(M/\msun)}-0.55 ,
\end{equation}
which has a linear slope of 0.38. The best fitting slope for our sample is $0.41\pm0.28$, thus agreeing with the value from \citet{mandushev91}. The large uncertainty arises from the scatter in Fig.~\ref{fig:mmlplane}.

The trend of increasing mass-to-light ratio with mass is expected from the models shown in Fig.~\ref{fig:model2} and the discussion in Sec.~\ref{sec:depend}. However, there the slope is $\sim0.6$---1.0 for metallicities $Z=0.0004$---0.004 and increases with $Z$. For some metallicities, the model slope is thus more than $1\sigma$ steeper than the fitted slope. This is not surprising, because the models each have a single dissolution timescale and metallicity, while in reality both quantities have a spread that causes horizontal and vertical scatter, respectively. It turns out that the spread in dissolution timescale has a stronger effect on $M/L$ than the spread in metallicity \citep{kruijssen08b}, implying that the scatter in the horizontal direction is largest and that the slope fitted to the entire sample is shallower than that of a single model.}

\begin{figure*}[t]
\resizebox{\hsize}{!}{\includegraphics[width=17cm]{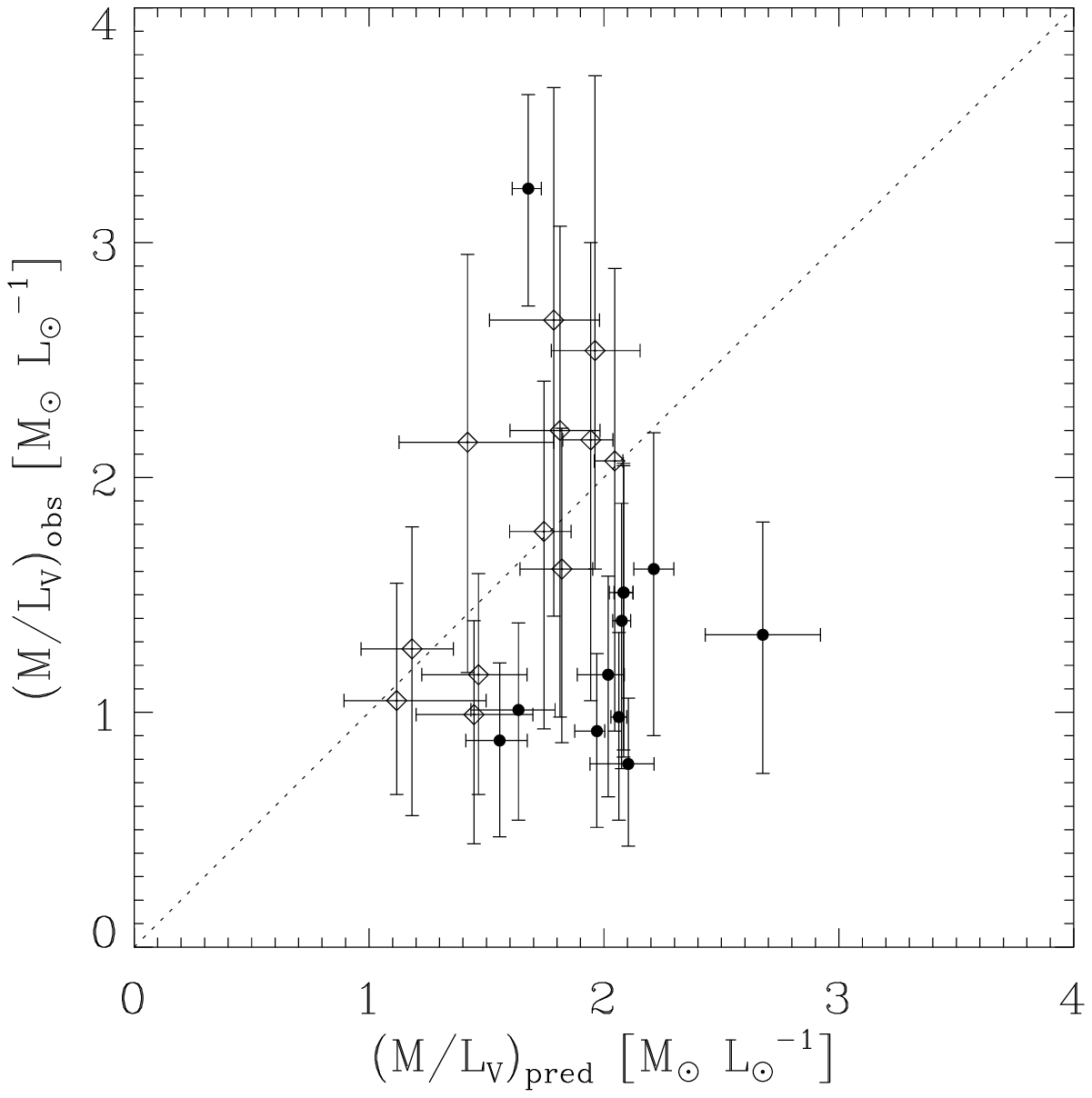}\includegraphics{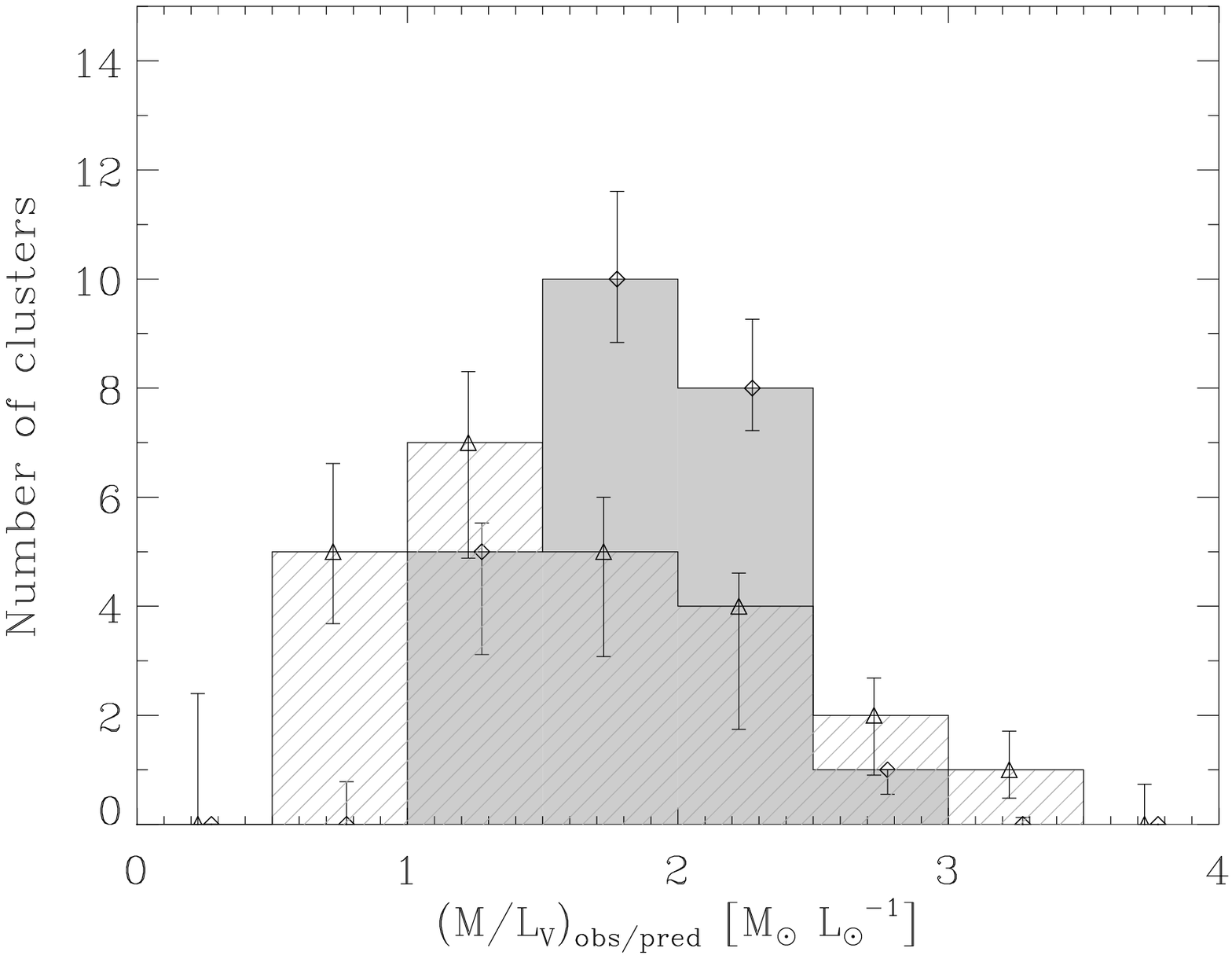}}
\caption[]{\label{fig:mlcomp}
      {\it Left}: Observed mass-to-light ratio $(M/L_V)_{\rm obs}$ versus the predicted mass-to-light ratio $(M/L_V)_{\rm pred}$, together with their $1\sigma$ standard errors. The dotted line follows the 1:1 relation. Clusters for which the disagreement is larger than 1$\sigma$ are plotted as dots. {\it Right}: Number histogram of $(M/L_V)_{\rm pred}$ (diamonds, shaded area) and $(M/L_V)_{\rm obs}$ (triangles, hashed area). Again, the error bars denote $1\sigma$ deviations, which were determined from 30k random realisations of the underlying data.
          }
\end{figure*}
In Fig.~\ref{fig:mlcomp}, a more specific comparison is made between the observations and model predictions using the same framework as for the canonical expectations in Fig.~\ref{fig:obs}. Again, the left-hand panel plots the observed versus the predicted mass-to-light ratios, while the right-hand panel shows the number histograms of the two. In the left-hand panel it is shown that the predictions for {half} of the clusters are such that they reach down to the appropriate mass-to-light ratios. When comparing this panel to its analog in Fig.~\ref{fig:obs}, the improved agreement with the observations is evident. Nonetheless, there is an aggregate of deviating GCs {\it below} the 1:1 relation at $(M/L_V)_{\rm pred}\approx 2~\msun~{\rm L}_\odot^{-1}$, representing the clusters for which no strong low-mass star depletion is expected from the models due to their long disruption times. Consequently, the predicted $M/L_V$ for these clusters are similar or equal to the canonical values. Except for NGC 6809, there are no clusters above the 1:1 relation that are inconsistent with the observations. The number histogram of the observed and predicted mass-to-light ratios in the right-hand panel of Fig.~\ref{fig:mlcomp} confirms both the improved agreement between observed and predicted $M/L_V$ with respect to Fig.~\ref{fig:obs} and the accumulation of a number of clusters near the canonical $M/L_V$ in the model predictions. 

\subsection{Discussion of discrepant clusters} \label{sec:unexpl}
{In total, there are twelve clusters with a worse than $1\sigma$
agreement between the model predictions and observations. Five of these have worse than $2\sigma$
agreement, while we would expect only one. Here, we discuss possible reasons behind the discrepancy.}

\begin{figure*}[t]
\resizebox{\hsize}{!}{\includegraphics[width=17cm]{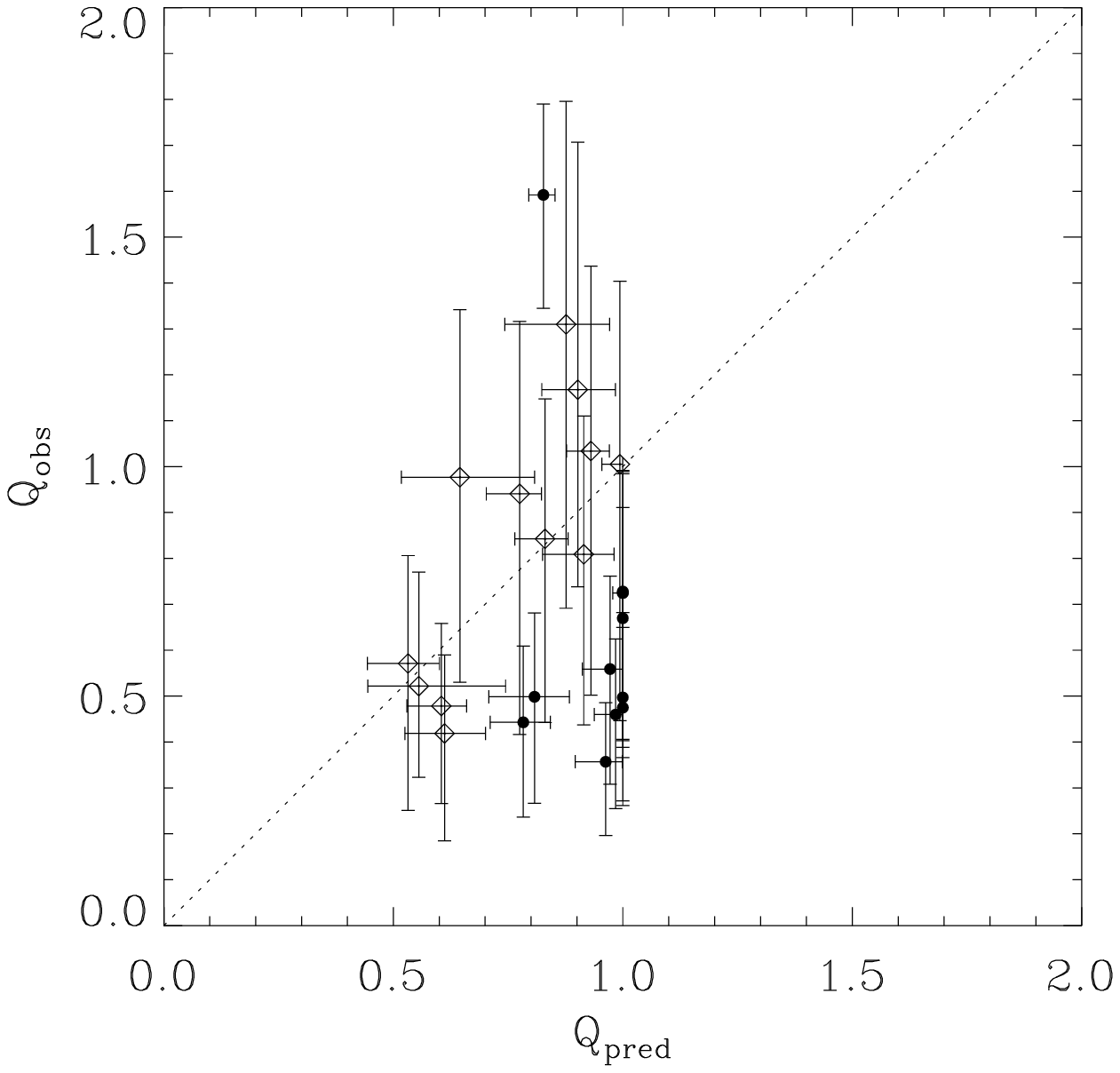}\includegraphics{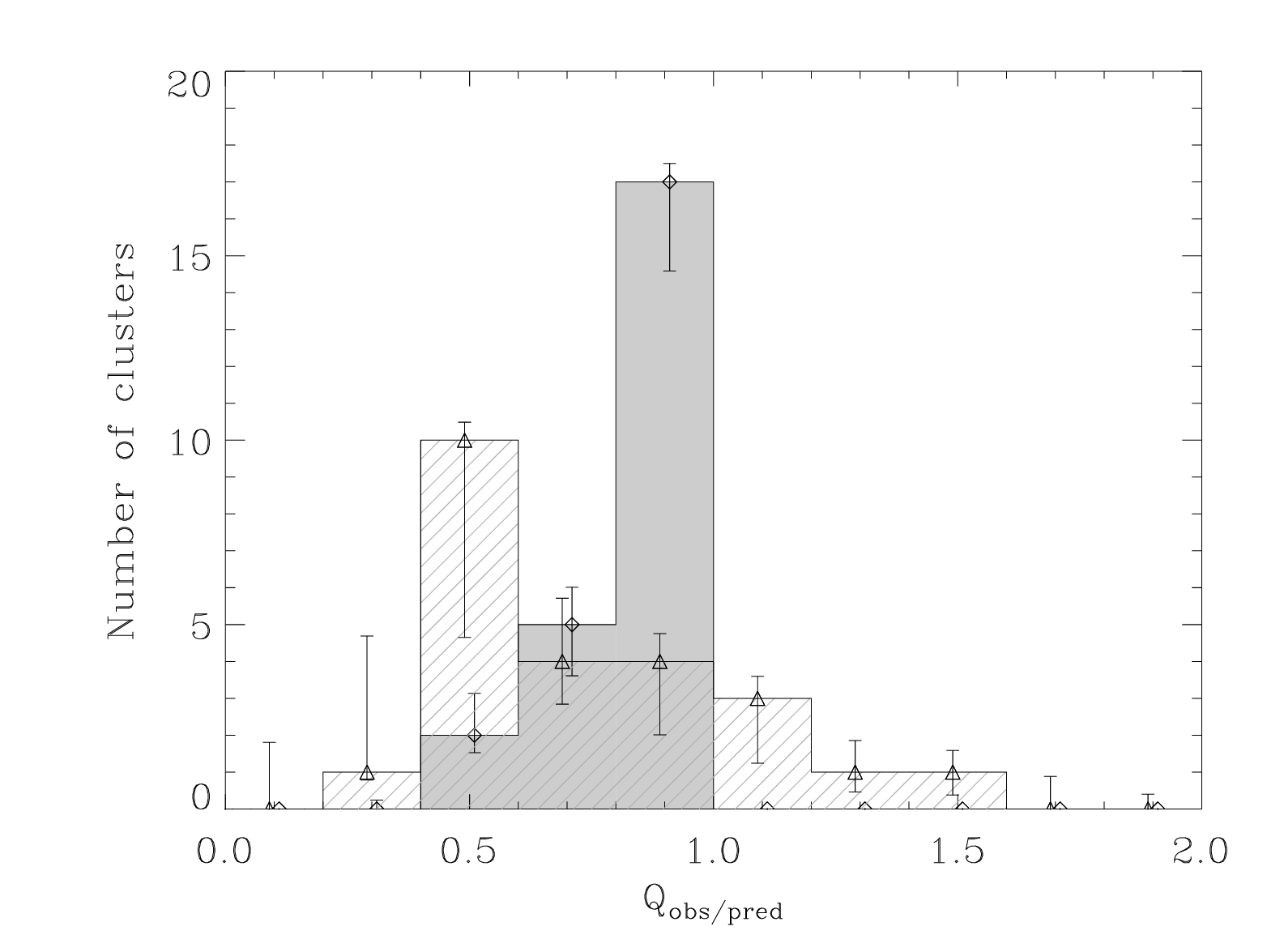}}
\caption[]{\label{fig:drop}
      {\it Left}: The ratio of the observed mass-to-light ratio $(M/L_V)_{\rm obs}$ to the canonically expected $(M/L_V)_{\rm can}$ versus the ratio of the predicted mass-to-light ratio $(M/L_V)_{\rm pred}$ to $(M/L_V)_{\rm can}$, together with their $1\sigma$ standard errors. The dotted line follows the 1:1 relation. Clusters for which the disagreement is larger than 1$\sigma$ are plotted as dots. {\it Right}: Number histogram of the ratio $\mathcal{Q}_{\rm pred}$ (diamonds, shaded area) and $\mathcal{Q}_{\rm obs}$ (triangles, hashed area). Again, the error bars denote $1\sigma$ deviations, which were determined from 30k random realisations of the underlying data.
          }
\end{figure*}

The deviant clusters below the 1:1 relation in the left-hand panel of Fig.~\ref{fig:mlcomp}, being NGC 104, 1851, 1904, 4147, 4590, 5272, 5904, 6205, 6341, 6934 and 7089, generally share properties such as relatively wide orbits and long dissolution timescales. Due to their long dissolution timescales, they are all predicted to have near-canonical $M/L_V$. This is illustrated in
Fig.~\ref{fig:drop}, where the fraction of $(M/L_V)_{\rm
obs}$ and $(M/L_V)_{\rm pred}$ with respect to the canonical
$(M/L_V)_{\rm can}$ is shown in panels similar to
Fig.~\ref{fig:mlcomp}. In both the left- and right-hand panels of
Fig.~\ref{fig:drop}, the accumulation of too high predicted
mass-to-light ratios occurs near or at $\mathcal{Q}_{\rm
pred}=1$. Since per definition $(M/L_V)_{\rm pred}\leq(M/L_V)_{\rm
can}$, no values $\mathcal{Q}_{\rm
pred}>1$ are found. In that range, the apparent
disagreement between the observed and predicted histograms is
disputable since all but one cluster (NGC 6809) are in $1\sigma$
agreement with their canonical mass-to-light ratios.

{While our predicted mass-to-light ratios are global (i.e., cluster-wide) values,} the observations from \citet{mclaughlin05} are derived from central velocity dispersion measurements from \citet{pryor93} and are extrapolated to global values using surface brightness profiles \citep[and references therein]{mclaughlin05}. They fit isotropic single-mass King models and thus neglect any radial gradients of $M/L$ ratio or mass function slope. Consequently, the values of $(M/L_V)_{\rm obs}$ do not contain any information about such gradients and for some clusters only accurately reflect the $M/L_V$ ratio in their central parts. The global $M/L_V$ ratios of clusters with strong radial $M/L$ gradients are at best {\it approximated} (McLaughlin, private communication). For instance, the centre of a mass-segregated cluster may be populated with massive, i.e., luminous stars, yielding a lower $M/L_V$ than its global value.

{The disagreement between the global and central $M/L$ is expected to be largest for clusters that have suffered relatively weak mass loss but are internally evolved. In that case, the low-mass stars are outside the core but still bound to the cluster and are included in the global $M/L$, while they do not play a role in the value derived by \citet{mclaughlin05}. This is indeed the case for the discrepant GCs in our sample, which not only have long dissolution timescales but also higher King parameters $W_0$, implying that mass segregation can be reached on relatively shorter timescales. For the GCs with worse than $1\sigma$ agreement below the 1:1 line in Figs.~\ref{fig:mlcomp} and~\ref{fig:drop} we have average King parameter $\overline{W_0}=7.6$, while for the $1\sigma$-consistent GCs we find $\overline{W_0}=6.1$, both with standard errors $<0.1$. This further validates our explanation for the difference between the central and global $M/L$ ratios of these GCs.

In a recent study, \citet{demarchi07} find that extended (low-concentration and low-$W_0$) GCs are depleted in low-mass stars, which they confirm to be in accordance with predictions by theoretical studies \citep{chernoff90,takahashi00}, while GCs with high values of $W_0$ have close to canonical mass functions. This is in agreement with our predicted $M/L$ for these clusters and suggests that the observed $M/L_V$ are indeed underestimated. A more precise check can be made by comparing the low-mass star depletion from \citet{demarchi07} with the observed and predicted fractions of the canonical $M/L_V$ ratios $\mathcal{Q_{\rm obs/pred}}$. This can be done for four GCs with worse than $1\sigma$ agreement, being NGC~104, 5272, 6341 and 6809. For NGC~104 and~5272 the observed depletion is not strong enough to draw any definitive conclusions, while for NGC~6341 and~6809 the results from \citet{demarchi07} are clearly more consistent with our predictions than with the observed $M/L_V$ (see also Sect.~\ref{sec:test} and Fig.~\ref{fig:alpha}). This substantiates the claim that some GCs have observed $M/L$ ratios that are biased to lower numbers. To test this assertion, {\it global} observational measurements of the velocity dispersion would be needed to enhance the accuracy of the present observed $M/L$ ratios.

We now revisit the mean $M/L_V$ fractions of the canonical value presented in Sect.~\ref{sec:mlcomp} by leaving out the GCs that may have strongly different central and global $M/L$ ratios. It was shown by \citet{baumgardt03} that for a $10^5~\msun$ cluster core collapse is reached within a Hubble time if $W_0\geq 7$. This timescale is increased by a factor three for a GC with typical initial mass of $10^6~\msun$, but mass segregation manifests itself on a shorter timescale than the core collapse time. The relative mass loss due to dissolution of a $10^6~\msun$ GC is smaller than 10\% after 12~Gyr for dissolution timescales $t_{0,{\rm tot}}\geq5$~Myr. These limits could separate GCs with similar global and central $M/L$ ratios from those with pronounced differences between the two. We exclude GCs with both $t_{0,{\rm tot}}\geq 5$~Myr {\it and} $W_0\geq7$, as well as NGC~6809 (which has a $M/L_V$ ratio that cannot be explained by any model as it is 1.6 times the canonical value). This yields an average observed fraction of the canonical $M/L_V$ of $\mathcal{Q}_{\rm obs}=0.78^{+0.09}_{-0.11}$ and a predicted value of $\mathcal{Q}_{\rm pred}=0.78\pm0.02$. For the excluded GCs, we have $\mathcal{Q}_{\rm obs}^{\rm excl}=0.68^{+0.05}_{-0.06}$ and $\mathcal{Q}_{\rm pred}^{\rm excl}=0.96\pm0.01$, reflecting the fundamental difference between both values. Although the cuts we made represent only `educated guesses', it is evident that the agreement between theory and observations is much better for those GCs for which we can be more certain that the central $M/L_V$ reflects the global value. For these GCs, our models confirm an average $M/L$ ratio drop of about 20\% due to low-mass star depletion, corresponding to about 1/4 of the observed difference in $M/L_V$ between GCs and UCDs.

Another option could be that the dissolution timescales of GCs on wide orbits are overestimated \citep[as suggested for different reasons by][]{kruijssen09b}, possibly due to a dissolution mechanism that is not included in our analysis. White dwarf kicks \citep{fregeau09} could be a candidate for such a mechanism. This would imply that some of our predicted dissolution timescales and $M/L$ ratios are overestimated. Also, we do not assume clusters to be initially mass-segregated. Some of the clusters under consideration here are likely not to have reached energy equipartition within a Hubble time, but still exhibit evidence for mass segregation \citep[e.g.,][]{anderson96}. This points to primordial mass segregation in these cases, which is shown by \citet{baumgardt08b} to effect additional low-mass star depletion that we did not account for (see also Sect.~\ref{sec:depend}). The additional modeled $M/L$ ratio decrease would be $\sim 10$\% (KL08). However, this is not sufficient to lift the discrepancy for any of the deviating GCs.}

\section{Observational verification} \label{sec:test}
If the decrease of $M/L$ ratio with respect to the canonical value is indeed due to low-mass star depletion, one would expect a correlation between the observed slope of the low-mass MF $\alpha_{\rm obs}$ and the ratio of the predicted and canonical $M/L_V$ ratios
$\mathcal{Q}_{\rm pred}$. Specifically, for a powerlaw MF with $n\propto m^{-\alpha}$, a small value of $\mathcal{Q}_{\rm pred}$ would be signified by a reduced value of $\alpha_{\rm obs}$. 

In a study by \citet{demarchi07}, MF slopes are
determined in the stellar mass range $m=0.3$---$0.8~\msun$ for
several Galactic globular clusters, based on a compilation of results
from HST imaging of different sources. By reanalysing the
\citet{baumgardt03} $N$-body data, \citet{baumgardt08b} conclude that
for a \citet{kroupa01} IMF the canonical slope in that mass range is
$\alpha_0=1.74$, which is thus expected to be measured for clusters
with canonical $M/L$ ratios or $\mathcal{Q}_{\rm pred}=1$. In
addition, they provide a fourth-order powerlaw fit to the $N$-body
simulations from \citet{baumgardt03} for $\alpha$ as a function of
$t/\tdis$, the elapsed fraction of the total disruption time. By
inverting our relation between $\mathcal{Q}_{\rm pred}$ and $t/\tdis$
(Eq.~\ref{eq:qtdis}) and inserting the outcome into $\alpha(t/\tdis)$
from \citet[Eq.~4]{baumgardt08b}, we obtain an expression for the
predicted MF slope $\alpha_{\rm pred}$ between 0.3 and 0.8~$\msun$ as
a function of the fraction of the canonical $M/L_V$ ratio $\mathcal{Q}_{\rm
pred}$. Inversion of Eq.~\ref{eq:qtdis} yields
\begin{equation}
  \label{eq:tdisq}
   t/\tdis = \left\{
  \begin{array}{ccl}
  1.25-0.85\mathcal{Q}_{\rm pred} & {\rm for} & \mathcal{Q}_{\rm pred}\leq 0.645, \\
  1.61-1.41\mathcal{Q}_{\rm pred} & {\rm for} & 0.645 < \mathcal{Q}_{\rm pred} < 1,  \\
  0...0.2 & {\rm for} & \mathcal{Q}_{\rm pred} = 1,  \\
  \end{array}\right.
\end{equation}
where the uncertainty $t/\tdis=0...0.2$ for $\mathcal{Q}_{\rm pred}=1$ arises due to the range of $t/\tdis$ over which it is constant {in our models}. Combination of this expression and Eq.~4 from \citet{baumgardt08b} then provides the relation between $\alpha_{\rm pred}$ and $\mathcal{Q}_{\rm pred}$:
\begin{equation}
  \label{eq:alphaq}
  \alpha_{\rm pred} = \left\{
  \begin{array}{ccccl}
   \displaystyle\sum_{n=0}^4 a_n\mathcal{Q}_{\rm pred}^n & {\rm for} & \mathcal{Q}_{\rm pred}\leq 0.645, \\
   \displaystyle\sum_{n=0}^4 b_n\mathcal{Q}_{\rm pred}^n & {\rm for} & 0.645 < \mathcal{Q}_{\rm pred} < 1,  \\
   1.68...1.74 & {\rm for} & \mathcal{Q}_{\rm pred} = 1,  \\
  \end{array}\right.
\end{equation}
with the coefficients $\{a,b\}_n$ listed in Table~\ref{tab:acoeff} and again the uncertainty $\alpha_{\rm pred}=1.68...1.74$ emerging from the degeneracy of $\mathcal{Q}_{\rm pred}=1$ that was mentioned earlier\footnote{Consequently, it represents the same uncertainty as $t/\tdis=0...0.2$, with $\alpha_{\rm pred}=1.68$ corresponding to $t/\tdis=0.2$ and $\alpha_{\rm pred}=1.74$ to $t/\tdis=0$. Since for most GCs under consideration the elapsed fractions of the total disruption time are closer to $t/\tdis=0.2$ than to $t/\tdis=0$ (see Table~\ref{tab:pred}), we adopt $\alpha_{\rm pred}=1.68$ if $\mathcal{Q}_{\rm pred}=1$.}.
\begin{table}[tb]\centering
\begin{tabular}{c |c c c c c}
  \hline \multicolumn{6}{c}{$\alpha_{\rm pred}$ coefficients} \\
  \hline \hline $n$ & 0 & 1 & 2 & 3 & 4 \\ \hline
     $a_n$ & -4.31 & 17.76 & -22.10 & 13.17 & -3.02    \\
     $b_n$ & -17.04 & 74.41 & -116.76 & 84.13 & -23.06  
                     \\\hline
\end{tabular}
\caption[]{\label{tab:acoeff}
     Coefficients for the fourth-order powerlaw approximation of $\alpha_{\rm pred}$ as a function of $\mathcal{Q}_{\rm pred}$ (see Eq.~\ref{eq:alphaq}).
    }
\end{table}

\begin{figure}[t]
\resizebox{\hsize}{!}{\includegraphics{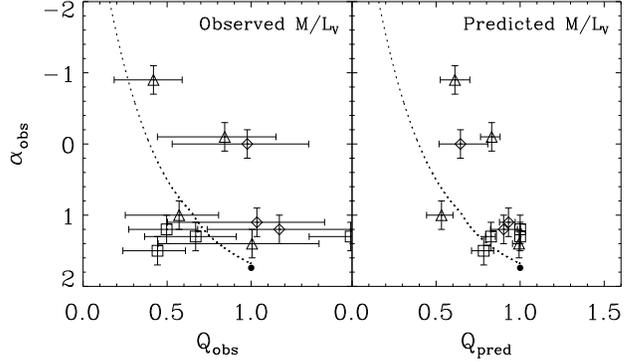}}
\caption[]{\label{fig:alpha}
      Correlation between the observed slope of the low-mass stellar mass function $\alpha_{\rm obs}$ and the relative mass-to-light ratio with respect to the canonical value $\mathcal{Q}$. The dotted curves indicate the theoretically predicted relation between $\alpha$ and $\mathcal{Q}$ (not a fit), with the dot at the right-hand tip representing the canonical values of $\alpha_0=1.74$ and $\mathcal{Q}=1$. {Values of $\alpha$ representing the {\it global} MF are marked with triangles, while those for clusters with worse than $1\sigma$ agreement between $\mathcal{Q}_{\rm obs}$ and $\mathcal{Q}_{\rm pred}$ are denoted by squares.} {\it Left}: For the observed mass-to-light ratio fraction $\mathcal{Q}_{\rm obs}$. {\it Right}: For the predicted mass-to-light ratio fraction $\mathcal{Q}_{\rm pred}$.
          }
\end{figure}
In Fig.~\ref{fig:alpha}, the correlation between $\mathcal{Q}_{\rm
obs/pred}$ and the observed low-mass MF slope $\alpha_{\rm obs}$ is
assessed for the subsample of clusters from the present study that is
also considered in \citet{demarchi07}. For comparison, the relation
for the predicted low-mass MF slope $\alpha_{\rm pred}$ as a function of $\mathcal{Q_{\rm obs/pred}}$
is included as well. Most of the observed data match the
predicted relation between $\alpha$ and $\mathcal{Q}$ within their error bars, albeit with substantial scatter. {This is due to the large uncertainties of the observations and possibly also related to biases introduced by comparing central and global mass-to-light ratios (see Sect.~\ref{sec:unexpl}). The poor quality of the observations is illustrated by this spread and by the large error bars. For the predicted mass-to-light ratios the trend is more well-defined, but for low values of $\alpha_{\rm obs}$ it does not extend down to the mass-to-light ratios that are predicted by theory. This could imply that either $\alpha_{\rm obs}$ or $\mathcal{Q_{\rm pred}}$ are biased. If the latter is true, it suggests that some GCs perhaps dissolve more rapidly than presently included in the models. 

As shown in Sect.~\ref{sec:unexpl}, comparison of $\alpha_{\rm obs}$ with the observed and predicted fractions of the canonical $M/L_V$ ratios $\mathcal{Q_{\rm obs/pred}}$ for the GCs with agreement parameter $\geq 2$ (see Table~\ref{tab:pred}) provides an independent check of our predicted $M/L_V$ ratios. While for NGC~104 and~5272 this does not allow for any definitive conclusions, for NGC~6341 and~6809 the observed mass functions are clearly more consistent with our predicted $M/L_V$ ratios than with the observed values.}

\begin{table}[tb]\centering
\begin{tabular}{c |c c}
  \hline \multicolumn{3}{c}{MF slopes} \\
  \hline \hline NGC & $\alpha_{\rm obs}$ & $\alpha_{\rm pred}$ \\ \hline
    104   &      $1.2\pm 0.3$   &    1.68$_{-0.00}^{+0.00}$                   \\
    288   &      $0.0\pm 0.3$   &    0.96$_{-0.31}^{+0.40}$              \\
  1851   &              &                       1.68$_{-0.00}^{+0.00}$                     \\
  1904   &              &                       1.65$_{-0.07}^{+0.03}$                      \\
  4147   &              &                       1.40$_{-0.18}^{+0.14}$                      \\
  4590   &              &                       1.66$_{-0.05}^{+0.02}$                     \\
  5139   &      $1.2\pm 0.3$   &    1.55$_{-0.11}^{+0.12}$                 \\
  5272   &      $1.3\pm 0.3$   &    1.68$_{-0.00}^{+0.00}$                       \\
  5466   &              &                       1.57$_{-0.12}^{+0.09}$                      \\
  5904   &              &                       1.64$_{-0.08}^{+0.04}$                     \\
  6093   &              &                       1.52$_{-0.20}^{+0.14}$               \\
  6121   &  ${\bf1.0\pm0.2}$ &   0.62$_{-0.32}^{+0.25}$                    \\
  6171   &              &                       1.34$_{-0.15}^{+0.10}$                  \\
  6205   &              &                       1.68$_{-0.00}^{+0.00}$                       \\
  6218   & $-{\bf0.1\pm0.2}$ &   1.44$_{-0.11}^{+0.09}$                    \\
  6254   &      $1.1\pm 0.3$   &    1.59$_{-0.07}^{+0.05}$                   \\
  6341   &      $1.5\pm 0.3$   &    1.35$_{-0.15}^{+0.12}$                   \\
  6362   &              &                       0.85$_{-0.21}^{+0.15}$                   \\
  6656   & ${\bf1.4\pm0.2}$  &   1.67$_{-0.04}^{+0.01}$                      \\
  6712   & $-{\bf0.9\pm0.2}$ &   0.87$_{-0.24}^{+0.24}$                   \\
  6779   &              &                       0.70$_{-0.37}^{+0.63}$                   \\
  6809   &      $1.3\pm 0.3$   &    1.44$_{-0.06}^{+0.04}$                    \\
  6934   &              &                       1.68$_{-0.02}^{+0.00}$                            \\
  7089   &              &                       1.68$_{-0.00}^{+0.00}$                            \\\hline
\end{tabular}
\caption[]{\label{tab:alpha}
     Observed low-mass stellar mass function (MF) slopes $\alpha_{\rm obs}$ in the range $m=0.3$---$0.8~\msun$ from \citet{demarchi07} (second column), and predicted values $\alpha_{\rm pred}$ for the same mass range based on the fit of Eq.~\ref{eq:alphaq} and Fig.~\ref{fig:alpha} (third column). The observed low-mass MF slope of clusters representing {\it global} MFs from multi-mass Michie-King models are denoted in boldface, while the other values designate local values close to the half-mass radius (see text). The standard errors on $\alpha_{\rm obs}$ is are $\sigma_\alpha=0.2$ for the global MFs and $\sigma_\alpha=0.3$ for the other values (De Marchi, private communication).
    }
\end{table}
With Eq.~\ref{eq:alphaq}, we can also {\it predict} the slope of the
low-mass MF for clusters that where not considered by
\citet{demarchi07}. The predicted slopes are listed in
Table~\ref{tab:alpha}. As stated above, for most clusters with observed values of $\alpha$, the agreement between observed and predicted $\alpha$
is reasonable. {Only for NGC~6218 and~6712 there is a strong
discrepancy. For NGC~6218, we expect the deviation to arise from
the observed value of $\alpha_{\rm obs}$, since the predicted and observed $M/L_V$ are in
excellent agreement (see Table~\ref{tab:pred}). On the other hand, for NGC~6712 the
incompatibility may be due to a slight overestimation of $(M/L_V)_{\rm
pred}$ and thus of $\mathcal{Q_{\rm pred}}$ and $\alpha_{\rm pred}$.}

\begin{figure*}[t]
\resizebox{8.6cm}{8.6cm}{\includegraphics{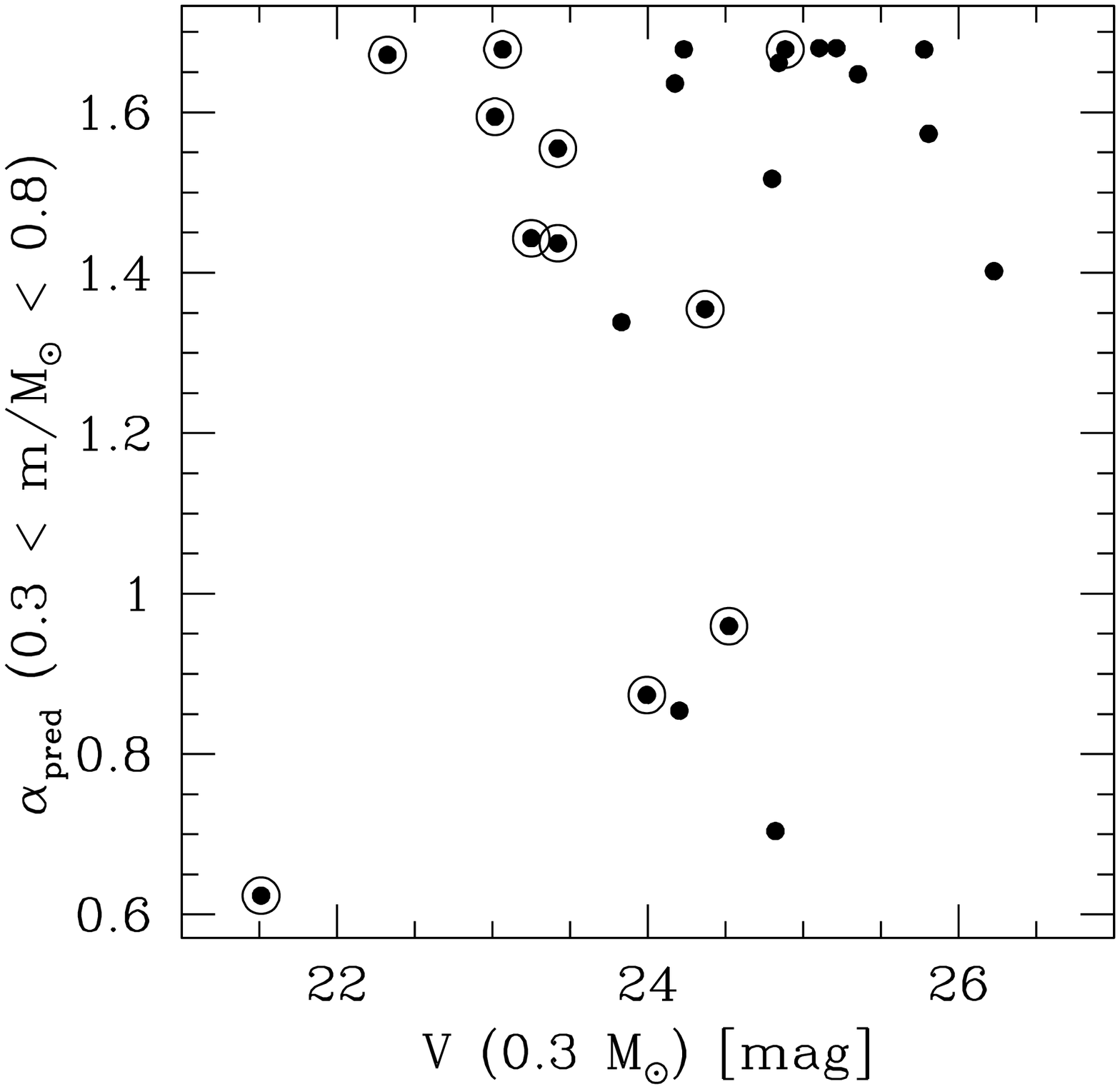}}
\resizebox{8.6cm}{8.6cm}{\includegraphics{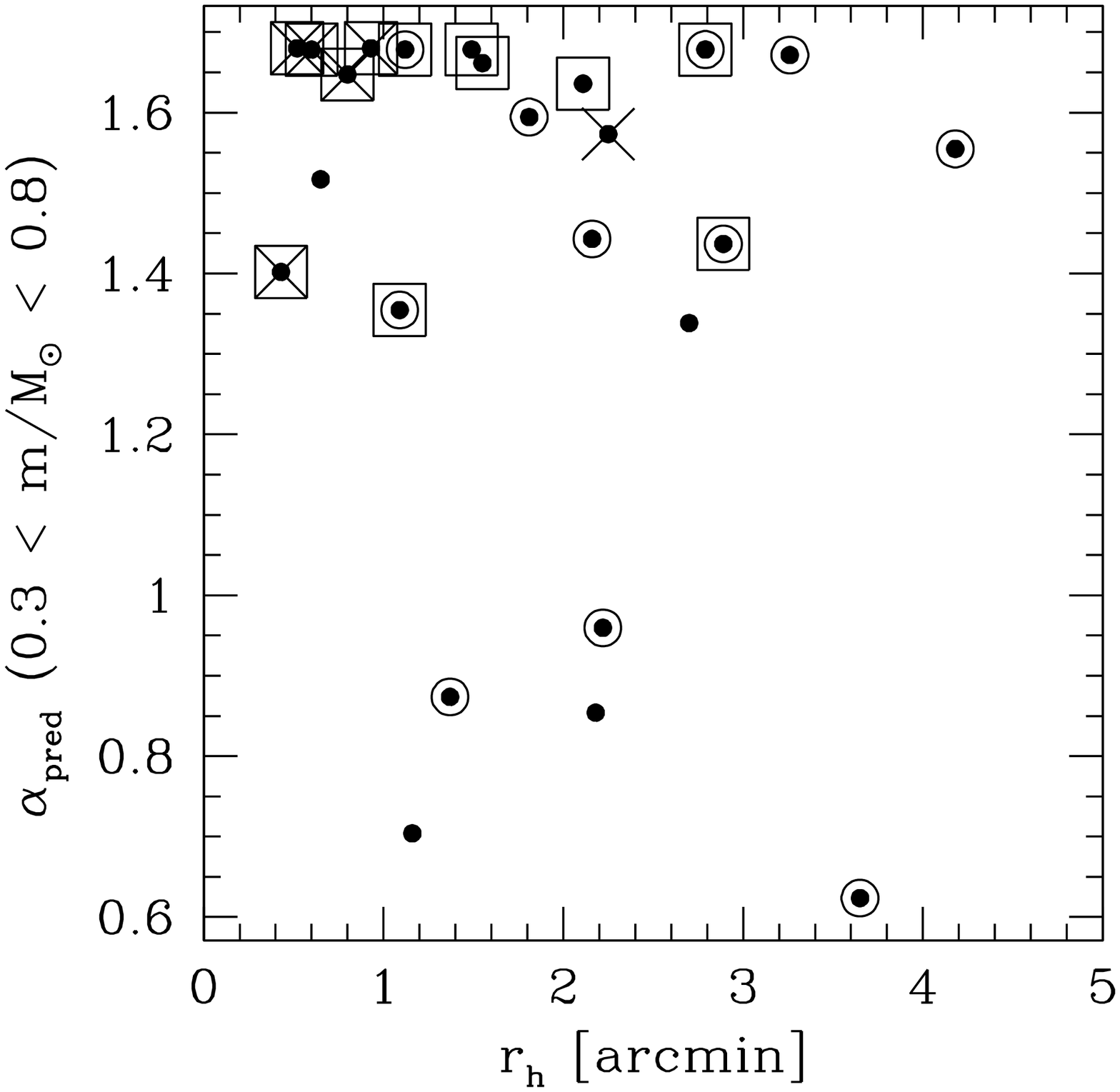}}
\caption[]{Feasibility of observational tests of low-mass star
      depletion. {\it Left}: For the 24 GCs investigated in this
      paper, the apparent $V$-band magnitude of stars with 0.3 solar masses
      $V_{0.3}$ is plotted versus the predicted slope $\alpha_{\rm
      pred}$ of the stellar mass function for
      $0.3<m/M_{\odot}<0.8$. Data points with large circles are those
      with available observational data from HST imaging (compiled by
      de Marchi et al. 2007). {\it Right}: The half-mass radius in
      arcminutes \citep{harris96} is plotted versus $\alpha_{\rm
      pred}$. Data points with crosses indicate GCs with $V_{0.3}>25$
      mag. Data points with large squares indicate GCs whose predicted
      $M/L_V$ deviates by more than 1$\sigma$ from the observed
      value. These are all GCs with agreement parameter $\ge 2$ in
      Table~\ref{tab:pred}. \label{fig:feasibility}}
\end{figure*}
In this context it must be noted that the compilation of $\alpha$
values from \citet{demarchi07} is drawn from a sample of literature
estimates, most based on HST data, observed in somewhat different
radial regions of each cluster. Four of the eleven GCs that coincide
with our sample of 24 GCs do have a direct estimate for their global
mass function (see Table~\ref{tab:alpha}). For the remaining seven other GCs from
\citet{demarchi07}, that estimate is taken from measurements
restricted to the region around the half-mass radius $r_{\rm h}$, {of which it is known that the shape of the MF is comparable to the global (i.e., cluster-wide) MF \citep{richer91,baumgardt03,demarchi07b}}. However, the uncertainty of these slopes is larger, and they do not provide a self-consistent way to derive the global MF. It is clear that a direct determination of the global stellar mass
function for most of the 24 GCs investigated in this study would allow
to verify the predictions of the present study with much higher
confidence.

In Fig.~\ref{fig:feasibility} we investigate how feasible it is to
      observationally verify the predicted drop of $\alpha$ for our
      full sample of 24 GCs. We plot the apparent $V$-band magnitude of stars
      with 0.3 solar masses $V_{0.3}$ for each cluster versus the
      predicted slope $\alpha_{\rm pred}$ of the stellar mass function
      for $0.3<m/M_{\odot}<0.8$. $V_{0.3}$ is obtained from the
      distance modulus of each GC and from the assumption that
      $M_{V_{0.3}}=9.8$ mag \citep{baraffe97}. We also plot the
      angular size of each cluster versus $\alpha_{\rm pred}$. As a
      consequence of their generally larger galactocentric distance,
      those GCs with the faintest $V_{0.3}>25$ mag would not be
      expected to exhibit a strong low-mass star depletion. The
      angular half-mass diameters of the GCs with $V_{0.3}<25$ range
      between 2 and 8 arcminutes. To obtain a representative estimate
      of the global mass function, it is clear that wide-field
      ground-based imaging is required for most GCs. For this
      wide-field imaging, a completeness magnitude of $V\sim$ 26 mag
      is desirable, which will allow moderately precise photometry
      already for $V\sim 25$ mag. For 8m class telescopes and with
      optical seeing in the range 0.8 to 1.0$''$, this requires 1-2
      hours integration time per filter, or 2-4 hours for a two-band
      exposure. With wide-field imagers such as VIMOS@VLT,
      IMACS@Magellan, or SuprimeCam@SUBARU, single-shot images will be
      sufficient to cover at least 2-3 half-light radii for most
      clusters. {From Fig.~\ref{fig:feasibility} we conclude that the best candidate that also complements the compilation by \citet{demarchi07} is NGC~6779, followed by NGC~6362 and possibly NGC~6171.}

\section{Discussion and conclusions} \label{sec:disc}
In this section, we provide a summary and a discussion of our results. We consider the effects of the assumptions that were
made and reflect on the implications of the results.

\subsection{Summary}
In this study, we have investigated the dynamical mass-to-light ratios
of 24 Galactic globular clusters. We have tested the hypothesis of the
preferential loss of low-mass stars as the main explanation for the
fact that the average observed mass-to-light ratios of the Galactic GCs {in our entire sample}
are only 74$^{+6}_{-7}$\% of the expectations from stellar population
models. Accounting for the orbital parameters we derived dissolution
timescales due to {two-body relaxation} and disc shocking for
our globular cluster sample and calculated the evolution of their
masses and photometry using the {\tt SPACE} analytical cluster models
from \citet[throughout this paper KL08]{kruijssen08}. These models
account for the preferential loss of low-mass stars {which is fitted to the $N$-body simulations by \citet{baumgardt03}} and therefore
provide non-canonical $M/L_V$ predictions. We find the derived dissolution timescales to be in good agreement with the range required for low-mass star depletion to explain the observed $M/L$ ratio decrease from \citet{kruijssen08b}.

{The present-day ($t=12$~Gyr) $M/L_V$ ratios have been compared to the observed values from \citet{mclaughlin05}, yielding $1\sigma$ agreement for
12 out of 24 GCs. We considered possible causes for the remaining $>1\sigma$ discrepancies that occur for the other GCs. It is found that 11 of these clusters have predicted $M/L_V$ very close to the canonically expected $M/L_V$ ratios due to their long dissolution timescales and the correspondingly modest low-mass star depletion, while their observed $M/L_V$ are lower. This is probably due to the method by which the observed $M/L_V$ are derived, which is biased towards the central $M/L_V$ while our models predict global $M/L_V$. For mass-segregated GCs with long dissolution timescales, both values can be substantially different. The discrepant GCs have higher than average King parameters $W_0$, which should indeed reach mass segregation on shorter timescales \citep[see e.g.,][]{baumgardt03}. This explanation for the discrepancy between some of the observed and predicted $M/L$ ratios is confirmed by a study of low-mass star depletion in GCs by \citet{demarchi07}, whose observed mass functions are in good agreement with our predictions. The average observed $M/L_V$ ratio of 74$^{+6}_{-7}$\% of the canonical expectations would therefore be underestimated. Excluding GCs which likely have dissimilar global and central $M/L$ ratios by making cuts in dissolution timescale and King parameter, we find that the observed and predicted $M/L_V$ ratios are consistent at 78$^{+9}_{-11}$\% and $78\pm2$\% of the canonical values, respectively. For the entire sample, the average predicted fraction of the canonical $M/L_V$ ratio is $85\pm1$\%.}

To assess the imprint of low-mass star depletion on the slope of the
low-mass stellar mass function, we compared the
observed mass function slopes $\alpha_{\rm obs}$ from \citet{demarchi07} for 11
GCs contained in our study to the values predicted by our models as well as to the observed and predicted mass-to-light ratio fractions of the canonical values $\mathcal{Q}_{\rm obs}$ and $\mathcal{Q}_{\rm pred}$. Most
of the measured slopes agree with the
predictions, but exhibit considerable scatter. Since most of them are
values derived at around the half-mass radius and are extrapolated to
global values, we also discuss the feasibility of observations for
{\it directly} measuring global mass functions of most of the GCs
investigated. We show that deep (ground-based) wide-field imaging
would be necessary, with point source detection limits $V\sim$26 mag. {The most suitable candidate for such a campaign would be NGC~6779.}

\subsection{Propagation of assumptions} \label{sec:assump}
In the course of the study presented in this paper, several assumptions were made that affect the results to different extents. Their implications are as follows.
\begin{itemize}
\item[(1)]
We have adopted the {\tt SPACE} cluster models (KL08), of which the stellar evolution and photometry are based on the Padova 1999 isochrones (see Sect.~\ref{sec:clevo}). Consequently, the predicted cluster photometry and corresponding mass-to-light ratios are affected by that choice. To indicate the level of the deviation with the cluster models from \citet{bruzual03}, in Fig.~\ref{fig:obs} we compared the canonically expected $M/L_V$ from {\tt SPACE} to the \citet{bruzual03} values for our cluster sample. The difference between both is inadequate to explain any systematic tendency of low mass-to-light ratio with respect to the {\tt SPACE} models. Therefore, we conclude that the adopted cluster models do not effect substantial implications for the predicted $M/L$ ratios\footnote{Of course, if the preferential loss of low-mass stars is accounted for, i.e., non-canonical models are considered, the {\tt SPACE} cluster models predict very different photometric evolution than canonical cluster models such as \citet{bruzual03}.}.
\item[(2)]
Due to the treatment of the preferential loss of low-mass stars in the {\tt SPACE} cluster models, there are indications that the predicted mass-to-light ratios could be underestimated during the final $\sim 15\%$ of the total cluster lifetime (KL08). Table~\ref{tab:pred} shows that none of the GCs in our sample reside in this regime. {On the other hand, we did not include primordial mass segregation, which could decrease the predicted $M/L$ ratios by $\sim 10$\% (KL08).}
\item[(3)]
By adopting the average cluster orbits from \citet{dinescu99}, we assume constant orbital parameters over the total cluster lifetimes. Considering the ballistic nature of the orbits, such an assumption is legitimate as long as the external conditions do not strongly differ. The Galactic potential was only substantially different from its present state during the formation of the Milky Way. A more extended distribution of mass during these early epoch would obviously {increase} the dissolution timescale due to disc shocking, and would affect the dissolution timescale due to {two-body relaxation} in a similar way because of the reduced tidal field. Consequently, this would imply that the mass loss during the first $\sim 1$~Gyr of our models is overestimated, causing our initial masses to be overestimated as well. {However, the extended nature of the Milky Way would cause dissolution due to giant molecular cloud encounters to become an important mechanism \citep[e.g.,][]{gieles06}, thereby counteracting the previous effect. Although we cannot rule out any consequences, a residual influence would only be relevant for a small fraction (the first $\sim 10$\%) of the total cluster lifetime, where mass loss by dissolution is much less effective than later on during cluster evolution. Therefore, this likely only affects our analysis within the error margins.}
\item[(4)]
We have compared our predictions to the observed mass-to-light ratios from \citet{mclaughlin05}, which are biased towards central $M/L$ values. As treated more extensively in Sect.~\ref{sec:unexpl}, this yields underestimated observed $M/L$ ratios for mass-segregated clusters with long dissolution timescales. Therefore, based on the earlier discussion and the parameter range in which discrepancies arise, we consider the dynamical $(M/L_V)_{\rm obs}$ from \citet{mclaughlin05} to be subject to improvement for GCs with both dissolution timescales $t_{0,{\rm tot}}\geq 5$~Myr {\it and} King parameters $W_0\geq 7$.
\end{itemize}

\subsection{Consequences and conclusions}
The consequences of our findings are not only relevant to studies of the $M/L$ ratios of compact stellar systems, but also to other properties of these structures. Here we list them together with the conclusions of this work.
\begin{itemize}
\item[(1)] 
{When constraining our sample to the subset for which the observed $M/L_V$ likely reflect the global values}, we find that the preferential loss of low-mass stars can account for the $\sim$20\% discrepancy between observed dynamical mass-to-light ratios of Galactic GCs and those expected from stellar population models that assume a canonical present day mass function \citep{kroupa01}. This alleviates the factor of two offset in $M/L$ between GCs and UCDs by about 25\%. Still, some additional dark mass with respect to a canonical IMF is required to explain the $M/L$ of most UCDs.
\item[(2)]
Accounting for the orbital parameters, present-day masses and chemical compositions of {\it individual} clusters, we find that there is good agreement between our model predictions and observations of the $M/L_V$ ratios of these clusters. {For the GCs with worse than $1\sigma$ agreement there are strong indications that the discrepancy is due to an underestimation of the observed $M/L$ ratio. In mass-segregated clusters with long dissolution timescales, the observed $M/L_V$ ratios represent central values that do not reflect the global $M/L_V$ ratio.}
\item[(3)]
The ideal way to confirm the validity of our explanation for the reduced $M/L$ ratios of GCs for {\it individual} clusters will be to obtain a homegeneous set of deep wide-field imaging for most GCs. This would expand and complement the currently available heterogeneous data sets of space based GC imaging, which is restricted to small fields in each GC, at different radial ranges. By this, the global mass and luminosity functions could be measured directly for individual GCs and be compared quantitatively to the predictions of this paper regarding the low-mass star depletion due to dynamical evolution. In addition, velocity dispersion measurements would allow for the determination of global $M/L$ ratios, thus providing an update to those from \citet{mclaughlin05}.
\item[(4)]
The topic of globular cluster self-enrichment and multiple stellar populations can also be considered within the framework of this paper. In a recent study by \citet{marino08} it is shown that NGC~6121 contains two stellar populations that are probably due to primordial variations in their respective chemical compositions. It is mentioned that the present-day mass of NGC~6121 is an order of magnitude smaller than that of known multiple-population GCs such as NGC~1851, 2808 and~5139. Consequently, \citet{marino08} pose the question how the enriched material could have remained in such a shallow potential and argue that multiple populations are unlikely to be strictly internal to GCs, unless they are the remnant of much larger structures. In the case of NGC~6121, our calculations seem to explain the issue, as it is the {\it initially fifth most massive GC} of our sample ($\mcli\sim 10^6~\msun$). As a result, mass could have been retained much more easily, implying that the multiple populations of NGC~6121 are no reason to invoke external processes for enrichment and to abandon the self-enrichment scenario.
\end{itemize}
We conclude that the variation of $M/L$ ratio due to cluster dissolution and low-mass star depletion is statistically significant and serves as a plausible explanation for the difference between observed and canonical $M/L$ ratios. Moreover, it has several implications that should be accounted for in GC studies, since its effects can be accurately quantified. We also suggest that the $M/L$ decrease is considered in independent observational verifications to further constrain the evolution of the stellar mass function in dissolving globular clusters.

\begin{acknowledgements}
We thank the anonymous referee for valuable comments that improved the manuscript. We are grateful to Dana Casetti-Dinescu for interesting discussions and for providing the disc shocking destruction rates from \citet{dinescu99}. Christine Allen is acknowledged for providing the data from \citet{allen06,allen08} in electronic form. We thank Guido De Marchi for sharing the uncertainties on the mass function slopes from \citet{demarchi07}. We particularly enjoyed stimulating discussions with Holger Baumgardt, Mark Gieles, Dean McLaughlin, Simon Portegies Zwart, Frank Verbunt and Reinier Zeldenrust. JMDK is grateful to Henny Lamers for support, advice and constructive comments on the manuscript, and is supported by a TopTalent fellowship from the Netherlands Organisation for ScientiÞc Research (NWO), grant number 021.001.038.
\end{acknowledgements}

\bibliographystyle{aa}
\bibliography{mybib}

\appendix
\section{Error analysis} \label{sec:app}
In this Appendix, the error propagation through our computations is discussed. The errors in Tables~\ref{tab:orbits} and~\ref{tab:t0} are standard errors, most of them determined by computing the formal error propagation. For a function $f(x_1,x_2,...,x_i)$ this implies
\begin{equation}
  \label{eq:error}
  \sigma_f^2=\left(\frac{\partial f}{\partial x_1}\right)^2\sigma_{x_1}^2+\left(\frac{\partial f}{\partial x_2}\right)^2\sigma_{x_2}^2+...+\left(\frac{\partial f}{\partial x_i}\right)^2\sigma_{x_i}^2 ,
\end{equation}
with $\sigma_i$ the error in the parameter $i$. Asymmetric errors on each parameter are both separately propagated by employing the same recipe, while inverse relations are accounted for by swapping the positive and negative errors. However, Eq.~\ref{eq:error} assumes an approximately constant derivative over the standard error interval. For very large errors on non-linear relations this assumption does not hold. The first of two parameters where we have to correct for this effect is $t_{0,{\rm sh}}$. It is inversely related to the destruction rate $\nu_{\rm sh}$ from \citet{dinescu99}, which is a parameter with very large relative errors, even to the extend that after computing the error propagation one can have $t_{0,{\rm sh}}-\sigma_{t_{0,{\rm sh}}}^-<0$. Because dissolution timescales smaller than zero are not physical, instead the negative error on $t_{0,{\rm sh}}$ is determined by computing 
\begin{eqnarray}
  \label{eq:t0sherror}
 \sigma_{t_{0,{\rm sh}}}^-&=&t_{0,{\rm sh}}(\nu_{\rm sh},\mc,x_{\rm corr}) \\
 \nonumber                        && -t_{0,{\rm sh}}(\nu_{\rm sh}-\sigma_{\nu_{\rm sh}}^-,\mc-\sigma_{M_{\rm cl}}^-,x_{\rm corr}-\sigma_{x_{\rm corr}}^-) ,
\end{eqnarray}
where $\sigma_i^-$ indicates the negative error in a parameter $i$. In the context of Eq.~\ref{eq:error}, this approach is equivalent to assuming the derivative equals the mean slope of $f(x)$ over the interval $[x-\sigma_x,x]$. On two occasions (NGC~6093 and~6712), a strongly asymmetric error in $t_{0,{\rm sh}}$ propagates into $t_{0,{\rm tot}}$ such that $t_{0,{\rm tot}}+\sigma_{t_{0,{\rm tot}}}^+\gg {\rm min}(t_{0,{\rm evap}}+\sigma_{t_{0,{\rm evap}}}^+,t_{0,{\rm sh}}+\sigma_{t_{0,{\rm sh}}}^+)$. However, since a very large positive error in $t_{0,{\rm sh}}$ or $t_{0,{\rm evap}}$ would make the term vanish in the inverse addition of Eq.~\ref{eq:t0}, it should not propagate into a similarly large error in $t_{0,{\rm tot}}$. This brings up the second parameter we have to correct for the propagation of large errors through non-linear relations. We define the error in $t_{0,{\rm tot}}$ for NGC~6093 and~6712 such that $t_{0,{\rm tot}}+\sigma_{t_{0,{\rm tot}}}^+=t_{0,{\rm evap}}+\sigma_{t_{0,{\rm evap}}}^+$.

The error margins on our predictions in Table~\ref{tab:pred} are determined by numerically evaluating Eq.~\ref{eq:error} for the desired quantities. Our predictions depend on the observed mass, metallicity and dissolution timescale. The derivatives of $M/L_V$ with respect to the former two are trivial since $M/L_V$ is determined by interpolating over these parameters. For the dissolution timescale, we compute additional models at $t_{0,{\rm tot}}-\sigma_{t_{0,{\rm tot}}}^-$ to obtain the numerical derivative of $M/L_V$ with respect to $t_{0,{\rm tot}}$. {In fact, this is the differential rather than the derivative, because for long dissolution timescales $M/L_V$ can be locally constant, while it varies over a larger range.} The only case were a non-linearity forces us to derive alternative errors is for the positive standard error on $M/L_V$. Although the uncertainty in metallicity could increase the predicted mass-to-light above its canonical value, the uncertainty in mass and dissolution timescale cannot due to the flattening of the cluster isochrones in the $\{M,M/L_V\}$-plane (see Fig.~\ref{fig:model2}). Therefore, the combined positive standard error of the mass-to-light ratio due to the uncertainty in mass and dissolution timescale $\sigma_{M/L_V}^{+,M,t_0}$ is defined as
\begin{equation}
  \label{eq:mlsigplus}
   \sigma_{M/L_V}^{+,M,t_0} = {\rm min}\left[\bar{\sigma}_{M/L_V}^{+,M,t_0},(M/L_V)_{\rm can}-(M/L_V)_{\rm pred}\right],
\end{equation}
with $\bar{\sigma}_{M/L_V}^{+,M,t_0}$ the standard error according to Eq.~\ref{eq:error}, $(M/L_V)_{\rm can}$ the canonically expected mass-to-light ratio and $(M/L_V)_{\rm pred}$ the predicted value. This definition ensures that the positive standard error is never larger than the difference between the canonical and predicted mass-to-light ratios.

Except for the alternative error in Eq.~\ref{eq:mlsigplus} {that is specific to $M/L_V$}, the standard errors on the predicted initial masses are determined analogously to the above. For the remaining lifetimes, numerical derivatives with respect to mass, metallicity and dissolution timescale are simply obtained by reintegrating Eq.~\ref{eq:dmdt} for slightly different initial conditions.

Finally, for the predicted slopes in Table~\ref{tab:alpha}, the errors are
computed using Eq.~\ref{eq:error} and restricted such that
$\alpha+\sigma_\alpha\leq\alpha_0$ (analogous to
Eq.~\ref{eq:mlsigplus}).

\end{document}